\documentclass{amsart}
\usepackage{amssymb}
\usepackage{amsmath}
\usepackage[T1]{fontenc}

\begin{document}
\newcommand{\ov}{\overline}
 \newcommand{\un}{\underline}
\renewcommand{\proof}{\bf {Proof:} \rm}

\newcommand{\BR}{{\mathbb R}}
\newcommand{\BC}{{\mathbb C}}
\newcommand{\BN}{{\mathbb N}}
\newcommand{\BZ}{{\mathbb Z}}

\newcommand{\m}{\underline{m}}
\renewcommand{\a}{\textbf{a}}
\renewcommand{\b}{\textbf{b}}
\newcommand{\p}{\underline{p}}
\newcommand{\q}{\mathfrak{q}}
\newcommand{\f}{\textbf{f}}
\newcommand{\edge}{\mathfrak{e}}
\newcommand{\g}{\textbf{g}}
\newcommand{\s}{\textbf{s}}
\newcommand{\G}{{\bf G}}

\newcommand{\e}{{\bf e}}
\newcommand{\vv}{{\bf v}}
\newcommand{\bb}{{\bf b}}
\newcommand{\C}{{\bf c}}

\newcommand{\ob}{{\bf b}}
\newcommand{\cW}{{W}}
\newcommand{\cL}{{L}}
\newcommand{\cC}{{C}}
\newcommand{\cM}{{M}}

\newcommand{\cl}{C \kern -0.1em \ell}

\renewcommand{\qed}{$\blacksquare$}
\newtheorem{theorem}{Theorem}[section]
\newtheorem{remark}{Remark}[section]
\newtheorem{lemma}{Lemma}[section]
\newtheorem{proposition}{Proposition}[section]
\newtheorem{corollary}{Corollary}[section]
\newtheorem{definition}{Definition}[section]
\newtheorem{example}{Example}[section]
\newtheorem{problem}{Problem}[section]

\title[Relativistic Wave Equations on the lattice]{Relativistic Wave Equations on the lattice: an operational perspective}

\author{N.~Faustino}
\address{CMCC, Universidade Federal do ABC, 09210--580, Santo Andr\'e, SP,
	Brazil
}
\email{nelson.faustino@ufabc.edu.br}
\dedicatory{Dedicated to Professor Wolfgang Spr\"o\ss ig on occasion of his $70th$ birthday.}
\date{\today}

\begin{abstract}  
This paper presents an operational framework for the computation of the discretized solutions for relativistic equations of Klein-Gordon and Dirac type. The proposed method relies on the construction of an evolution-type operador from the knowledge of the \textit{Exponential Generating Function} (EGF), carrying a \textit{degree lowering operator} $L_t=L(\partial_t)$. 
We also use certain operational properties of the discrete Fourier transform over the $n-$dimensional \textit{Brioullin zone}  $Q_h=\left(-\frac{\pi}{h},\frac{\pi}{h}\right]^n$ -- a toroidal Fourier transform in disguise -- to describe the discrete counterparts of the \textit{continuum} wave propagators, $\cosh(t\sqrt{\Delta-m^2})$ and $\dfrac{\sinh(t\sqrt{\Delta-m^2})}{\sqrt{\Delta-m^2}}$ respectively, as discrete convolution operators.
 In this way, a huge class of discretized time-evolution problems of differential-difference and difference-difference type may be studied in the spirit of hypercomplex variables.
\end{abstract}



\maketitle

\section{Introduction}

\subsection{The state of art}
Some  decades ago K.~G\"urlebeck and W. Spr\"ossig have shown on their book \cite{GuerlebeckSproessig97} that the theory of finite difference potentials offers the possibility to study the solution of boundary value problems from the knowledge of the discrete fundamental solution, when a finite difference approximation of the Dirac operator is considered. Such method was sucessfully applied on the papers \cite{FaustinoGHK06,CFaustinoV08} to compute numerically the solution of boundary value problems.
 
There are already some recent contributions that recognizes that the theory of finite difference potentials presented by G\"urlebeck and Spr\"ossig on their book may also be used to describe the solution of boundary value problems on half-lattices \cite{CKKS14,CKK15} by a discrete version of the Hilbert transform. 

It is almost well-known that the [discrete] Hilbert transform is nothing else than a Riesz type transform in disguise (cf.~\cite{Bersntein16,Bersntein17}), that may be derived formally from the subordination formula
$$\left(-\Delta_h\right)^{-\alpha}=\frac{1}{\Gamma(\alpha)}\int_0^{\infty} \exp\left(t\Delta_h\right)t^\alpha \frac{dt}{t}$$
in the limit $\alpha \rightarrow \frac{1}{2}$ (cf.~\cite[section 6.]{CiaurriGRTV17}).
Hereby $\Delta_h$ denotes the star-Laplacian operator on the lattice $h\BZ^n$ (that will be introduced later on subsection \ref{outlinePaper} via equation (\ref{discreteLaplacian})).
Thereby, the solutions of Riemann-Hilbert type problems may be recovered by the solution of a time-evolution problem of Cauchy-Riemann type (cf.~\cite[Proposition 6.]{CiaurriGRTV17}).

A great deal of work has been done recently by Dattoli and his collaborators to extend the operational framework to relativistic wave equations of Dirac type (cf.~\cite{BDattoliQ11,DattoliSGHP15,DattolliT15,DattoliGHPS17}). At the same time
there has been interest in studying evolution problems in the context of hypercomplex variables, namely discretized variants for the heat equation (cf.~\cite{BaaskeBRS14}) and for the Cauchy-Kovaleskaya extension (cf.~\cite{ConstalesDeRidder14}).

\subsection{Problem setup}\label{outlinePaper}

Let $\e_1,\e_2,\ldots,\e_n$, 
$\e_{n+1},\e_{n+2}\,\ldots,\e_{2n}$ be an orthogonal basis of the Minkowski space-time $\BR^{n,n}$, and $\cl_{n,n}$ the Clifford algebra of signature $(n,n)$ generated from the set of
graded anti-commuting relations
\begin{eqnarray}
\label{CliffordBasis}
\begin{array}{lll}
\e_j \e_k+ \e_k \e_j=-2\delta_{jk}, & 1\leq j,k\leq n \\
\e_{j} \e_{n+k}+ \e_{n+k} \e_{j}=0, & 1\leq j,k\leq n\\
\e_{n+j} \e_{n+k}+ \e_{n+k} \e_{n+j}=2\delta_{jk}, & 1\leq j,k\leq
n.
\end{array}
\end{eqnarray}

Here we recall that the linear space isomorphism provided by the linear extension of the mapping
$\e_{j_1}\e_{j_2}\ldots \e_{j_r} \mapsto dx_{j_1}dx_{j_2}\ldots
dx_{j_r}$, with $1\leq j_1<j_2<\ldots<j_r\leq 2n$, allows us to show that the resulting
algebra has dimension $2^{2n}$ and it is isomorphic to the exterior algebra
$\bigwedge (\BR^{n,n})$ (cf.~\cite[Chapter 3]{VazRoldao16}) so that $\e_{J}=\e_{j_1}\e_{j_2}\ldots
\e_{j_r}$ corresponds to a basis of $\cl_{n,n}$. For $J=\varnothing$ (empty set)
we use the convention $\e_\varnothing=1$.
In particular, any vector $(x_1,x_2\ldots,x_n)$ of $\BR^n$ may be represented in terms of the linear combination $\displaystyle x=\sum_{j=1}^n x_j \e_j$ carrying the basis elements $\e_1,\e_2,\ldots,\e_n$ with signature $(0,n)$, whereas the translations $(x_1,x_2,\ldots, x_j\pm \varepsilon,\ldots,x_n)$
on the lattice $\varepsilon\BZ^n \subset \BR^n$
with mesh width $\varepsilon>0$ may be represented in terms of the displacements $x\pm \varepsilon\e_j$.

Along this paper we develop our results to lattices of the form
\begin{eqnarray*}
\BR^n_{h,\alpha}:=(1-\alpha) h\BZ^n \oplus \alpha h\BZ^n,&\mbox{with }~~~~h>0&\mbox{and}~~~~0< \alpha< \frac{1}{2}. 
\end{eqnarray*}

Here we would like to stress that $\BR^n_{h,\alpha}$ contains $h\BZ^n$, since any $x$ with membership in $h\BZ^n$ may be rewritten as $(1-\alpha)x+\alpha x$, with $(1-\alpha) x \in (1-\alpha)h\BZ^n$ and $\alpha x \in \alpha h\BZ^n$.

The class of multi-vector functions $\BR^n_{\alpha,h} \rightarrow \BC\otimes\cl_{n,n}$ and $\BR^n_{\alpha,h} \times T \rightarrow \BC\otimes\cl_{n,n}$ that are considered on the sequel admit one of the following representations:
\begin{eqnarray*}
	\Phi(x)=&\displaystyle \sum_{r=0}^n\sum_{|J|=r} \phi_J(x) \e_J, &
	\mbox{with}~\e_{J}=\e_{j_1}\e_{j_2}\ldots \e_{j_r} \\
	\Psi(x,t)=&\displaystyle \sum_{r=0}^n\sum_{|J|=r} \psi_J(x,t) \e_J, &
	\mbox{with}~\e_{J}=\e_{j_1}\e_{j_2}\ldots \e_{j_r}.
\end{eqnarray*}

Hereby $|J|$ denotes the cardinality of $J$. 
The scalar-valued functions $\Phi(x)$ resp. $\Psi(x,t)$ are thus represented as $\Phi(x)=\phi(x) \e_\emptyset$ resp. $\Psi(x,t)=\psi(x,t) \e_\emptyset$,  whereas the vector-fields $(\phi_1(x),\phi_2(x),\ldots,\phi_n(x))$ and $(\psi_1(x,t),\psi_2(x,t),\ldots,\psi_n(x,t))$ of $\BR^n$ are described through the ansatz $
\displaystyle \Phi(x)=\sum_{j=1}^n \phi_j(x)\e_j
$ and
$
\displaystyle \Psi(x,t)=\sum_{j=1}^n \psi_j(x,t)\e_j
$, respectively.

The subscript notations $\phi_J(x)$ and $\psi_J(x,t)$ are adopted to denote the complex-valued functions $\BR^n_{h,\alpha}\rightarrow \BC$ resp. $\BR^n_{h,\alpha} \times T \rightarrow \BC$ carrying the multivector basis $\e_J$. The bold notations $\f,\g,\ldots,\Phi,\Psi,\ldots$ and so on will be considered when we refer to multivector functions with membership in the \textit{complexified Clifford algebra} $\BC\otimes\cl_{n,n}$.

Our purpose here is centered around the study of relativistic wave equations of Klein-Gordon and Dirac-type on the space-time lattice $\BR_{h,\alpha}^n\times T$ that exhibit a differential-difference or a difference-difference character. That includes time-evolution problems encoded by the discretized Klein-Gordon operator $L_t^2-\Delta_h+m^2$, carrying the mass term $m>0$. 

Here and elsewhere 
\begin{eqnarray}
\label{discreteLaplacian}
\displaystyle \Delta_h \Psi(x,t)=\sum_{j=1}^n
\frac{\Psi(x+h\e_j,t)+\Psi(x-h\e_j,t)-2\Psi(x,t)}{h^2}.
\end{eqnarray}
denotes the discrete Laplacian on $h\BZ^n\subset \BR_{h,\alpha}^n$, and $L_t$ a \textit{degree-lowering operator}.

The Dirac-K\"ahler discretizations $D_\varepsilon$ on the lattice $\varepsilon\BZ^n$, already studied in the author's recent papers \cite{FaustinoKGordonDirac16,FaustinoMMAS17}: 
\begin{eqnarray}
\label{DiracEqh}
\begin{array}{lll}
D_\varepsilon \Psi(x,t)&=&\displaystyle \sum_{j=1}^n\e_j\frac{\Psi(x+\varepsilon \e_j,t)-\Psi(x-\varepsilon\e_j,t)}{2\varepsilon}+ \\
&+&	\displaystyle \sum_{j=1}^n\e_{n+j}\frac{2\Psi(x,t)-\Psi(x+\varepsilon \e_j,t)-\Psi(x-\varepsilon \e_j,t)}{2\varepsilon}
\end{array}
\end{eqnarray}
as well as the pseudo-scalar $\gamma$ of $\cl_{n,n}$:
\begin{eqnarray}
\label{PseudoScalar}\gamma=\prod_{j=1}^n \e_{n+j}\e_j.
\end{eqnarray}
are also considered with the aim of formulate a discrete counterpart to the time-evolution equation of Dirac type.

From now on let us take a close look for the \textit{delta operators} $L_t$ from an umbral calculus perspective (see e.g. \cite[Chapter 1]{Faustino09}, \cite[section 1.]{Faustino10}, \cite[subsection 2.1.]{FaustinoR11} and \cite[subsection 1.2. \& subsection 2.2.]{FaustinoMonomiality14} for an abridged version of Roman's book \cite{Roman84}).

In case where $L_t=\partial_t$ and $T=[0,\infty)$, it is well-known (and easy to check) that the hypergeometric series representation of the following wave propagators (cf.~\cite[p.~704]{DattolliT15}):
\begin{eqnarray*}
	\cosh(t\sqrt{\Delta_h-m^2})&=& {~}_0F_1\left(\frac{1}{2};\frac{t^2}{4}(\Delta_h-m^2)\right) \\
	\dfrac{\sinh(t\sqrt{\Delta_h-m^2})}{\sqrt{\Delta_h-m^2}}&=& t{~}_0F_1\left(\frac{3}{2};\frac{t^2}{4}(\Delta_h-m^2)\right)
\end{eqnarray*}
allows us to represent formally the null solutions of the differential-difference Klein-Gordon operator $\partial_t^2-\Delta_h+m^2$ (cf.~\cite[Part I]{Dattoli97} \& \cite[\textsc{Exercise 2.18}]{Tao06}), while for the difference-difference evolution problem associated to the discretization $T=\{\frac{k\tau}{2}~:~k\in \mathbb{N}_0\}$ of the continuous time-domain $[0,\infty)$ (the lattice $\frac{\tau}{2}\BZ_{\geq 0}$), and to the finite difference operator
\begin{eqnarray}
\label{DifferenceDifferenceLt}L_t \Psi(x,t)=\dfrac{\Psi\left(x,t+\frac{\tau}{2}\right)-\Psi\left(x,t-\frac{\tau}{2}\right)}{\tau}
\end{eqnarray}
(difference-difference evolution problem) it can be easily verified that $L_t$ admits the \textit{formal Taylor series expansion} (cf.~\cite[Example 2.3.]{FaustinoMonomiality14}) $$L_t\Psi(x,t)=\frac{2}{\tau}\sinh\left(\frac{\tau}{2}\partial_t\right)\Psi(x,t).$$ 

Using the fact that first order differential and difference operators are particular cases of \textit{shift-invariant operators} with respect to the exponentiation operator $\exp\left(s \partial_t\right)$: $$L_t\exp\left(s \partial_t\right)=\exp\left(s \partial_t\right)L_t,$$
we can obtain an amalgamation of our approach to \textit{delta operators} $L_t$, represented through the formal series expansion
$$ 	L_t=\sum_{k=1}^\infty b_k \dfrac{\left(\partial_t\right)^k}{k!},~~~\mbox{with}~~~b_k=[(L_t)^k t^k]_{t=0},$$
in the same order of ideas of \cite[\textsc{Chapter 1}]{Faustino09}, \cite[section 2]{FaustinoR11} \& \cite[section 1 \& section 2]{FaustinoMonomiality14}.
Indeed, from the combination of the \textit{shift-invariant property} (cf.
\cite[Corollary 2.2.8]{Roman84}) with the isomorphism between the
algebra of formal power series and the algebra of linear
functionals associated to the ring of polynomials $\BR[t]$ (cf. \cite[Theorem 2.1.1]{Roman84}), the null solutions of the wave-type operator $L_t^2-\Delta_h+m^2$ may be constructed from \textit{exponential generating function} (EGF)
\begin{eqnarray}
\label{EGFLt}
\G(\s,t)=\sum_{k=0}^\infty \frac{m_k(t)}{k!}\s^k, & \s \in \BC\otimes\cl_{n,n}&\&~~~~t\in \BR
\end{eqnarray}
associated to the Sheffer sequence $\{m_k(t)~:~k\in \BN_0\}$ of $L_t:=L(\partial_t)$.

More precisely, the following theorem (see \textsc{Appendix} \ref{EGFsection}. for further details) goes beyond the Pauli matrices identity obtained in \cite[p.~701]{DattolliT15}:
\begin{theorem}\label{EGFhypercomplex}
	For the case where $\s=re^{i\phi}\omega$, with $-\pi<\phi\leq \pi$ and $\omega$ is an element of $\cl_{n,n}$ satisfying $\omega^2=+1$, the exponential generating function $\G(\s,t)$ defined through equation (\ref{EGFLt}), satisfies 
	$$ \G(re^{i\phi}\omega,t)=\cosh\left(tL^{-1}(re^{i\phi})\right)+\omega\sinh\left(tL^{-1}(re^{i\phi})\right)~.$$
\end{theorem}

\subsection{The structure of the paper}
We turn next with the outline of the subsequent sections:
\begin{itemize}
\item In section \ref{Setting} we introduce, in a self-contained style, the basics of discrete Fourier analysis on the lattice $\BR_{h,\alpha}^n$ (subsection \ref{DiscreteFourierSetting}). Then we obtain an alternative factorization for the discretized Klein-Gordon operator (subsection \ref{DiracLaplaceSection}), based on the study of the Fourier multiplier underlying to the discrete Laplacian $\Delta_h$ defined by eq. (\ref{discreteLaplacian}).
\item In section \ref{discretizedTimeEvolutions} we find some explicit representations underlying to the solution of the discretized versions of the Klein-Gordon (\textbf{Theorem \ref{mainResultdiffKleinGordon}}) and Dirac equation (\textbf{Corollary \ref{mainResultdiffDirac}}) on the lattice  $\BR_{h,\alpha}^{n} \times T$. Here, the EGF representation obtained in \textbf{Theorem \ref{EGFhypercomplex}} as well as the discrete Fourier analysis toolbox introduced in section \ref{Setting} play a central role. 
\item In section \ref{FurtherApplications} we study applications and generalizations for the results obtained in section \ref{discretizedTimeEvolutions}. 
 We start to find explicit representations for difference-difference evolution problems of Klein-Gordon and Dirac type on the lattice $\BR_{h,\alpha}^n \times \frac{\tau}{2}\BZ_{\geq 0}$ by means of hypersingular integral representations involving Chebyshev polynomials of first and second kind (cf.~\cite{MasonChebyshev93}), and of fractional integral representations associated to a class of \textit{generalized Mittag-Leffler functions} (cf. \cite[Chapter 1]{SamkoEtAl93}). We also establish a comparison with the approaches considered in references \cite{ConstalesDeRidder14,BaaskeBRS14,CiaurriGRTV17} (subsection \ref{discreteHeatsemigroup}). In the end, we exploit the characterization obtained in section \ref{discretizedTimeEvolutions} to fractional operators of Riesz type (subsection \ref{DiscreteFractionalCalculus}). 
\item In section \ref{Outlook} we outline the main results of the paper and discuss further directions of research. 
\end{itemize}

\section{Discrete Fourier Analysis toolbox}\label{Setting}

\subsection{Discrete Fourier transform vs. spaces of tempered distributions}\label{DiscreteFourierSetting}

Let us define by $\ell_p(\BR^n_{h,\alpha};\BC\otimes \cl_{n,n}):=\ell_p(\BR^n_{h,\alpha})\otimes \left(\BC\otimes \cl_{n,n}\right)$ ($1\leq p\leq \infty$) the \textit{right Banach-module} endowed by the Clifford-valued sesquilinear form (cf.~\cite[p.~533]{FaustinoBayesian17})
\begin{eqnarray}
	\label{lpInner} \langle \f(\cdot,t),\g(\cdot,t) \rangle_{h,\alpha}=\sum_{x\in \BR^n_{h,\alpha}}h^n~
	\f(x,t)^\dag\g(x,t),
\end{eqnarray}
and by $\mathcal{S}(\BR^n_{h,\alpha};\cl_{n,n}):=\mathcal{S}(\BR^n_{h,\alpha})\otimes \left(\BC\otimes \cl_{n,n}\right)$ the space of \textit{rapidly decaying functions} $\f(\cdot,t)$ ($t\in T$ is fixed) with values on $\BC\otimes\cl_{n,n}$, defined through the semi-norm condition
$$ \displaystyle \sup_{x \in \BR^n_{h,\alpha}} (1+\| x\|^2)^M~\| \f(x,t)\|<\infty$$
for any $\BR-$valued constant $M<\infty$.

Here and elsewhere, the symbol $\dag$ denotes the $\dag-${\it
	conjugation} operation $\a \mapsto\a^\dag$ on the \textit{complexified Clifford algebra} $\BC\otimes\cl_{n,n}$, defined as 
\begin{eqnarray}
\label{dagconjugation}
\begin{array}{lll}
(\a \b)^\dag=\b^\dag\a^\dag \\ (a \e_J)^\dag =\overline{a_J}~\e_{j_r}^\dag
\ldots \e_{j_2}^\dag\e_{j_1}^\dag~~~(1\leq j_1<j_2<\ldots<j_r\leq 2n) \\
\e_j^\dag=-\e_j~~~\mbox{and}~~~\e_{n+j}^\dag=\e_{n+j}~~~(1\leq j\leq
n)
\end{array},
\end{eqnarray}
whereas $\| \cdot \|$ -- the norm of the \textit{complexified Clifford algebra} $\BC\otimes \cl_{n,n}$ -- is defined by the square condition $\| \a\|^2=\a^\dagger \a$. 

In the same order of ideas of \cite[Exercise 3.1.7]{RuzhanskyT10}, under the seminorm constraint
$$ \displaystyle \sup_{x \in \BR^n_{h,\alpha}} (1+\| x\|^2)^{-M}~\| \g(x,t)\|<\infty$$
the mapping $\f(\cdot,t) \mapsto \langle \f(\cdot,t),\g(\cdot,t)\rangle_{h,\alpha}$ defines the set of all \textit{continuous linear functionals} with membership in $\mathcal{S}(\BR^n_{h,\alpha};\BC\otimes \cl_{n,n})$. The underlying family of distributions $\g(\cdot,t):\BR^n_{h,\alpha} \rightarrow \BC\otimes \cl_{n,n}$ (for a fixed $t\in T$) belong to $$\mathcal{S}'(\BR^n_{h,\alpha};\BC\otimes \cl_{n,n}):=\mathcal{S}'(\BR^n_{h,\alpha})\otimes\left(\BC\otimes \cl_{n,n}\right),$$ the multivector counterpart of the \textit{space of tempered distributions} on the lattice $\BR^n_{h,\alpha}$. 
Let us now take a close look to the \textit{discrete Fourier transform}, defined as follows:
\begin{eqnarray}
\label{discreteFh}
	(\mathcal{F}_{h,\alpha} \g)(\xi,t)&=&\left\{\begin{array}{lll}
		\displaystyle \frac{h^n}{\left(2\pi\right)^{\frac{n}{2}}}\displaystyle 
		\sum_{x\in \BR^n_{h,\alpha}}\g(x,t)e^{i x \cdot \xi} & \mbox{for} & \xi\in Q_h
		\\ \ \\
		0 & \mbox{for} & \xi\in \BR^n \setminus Q_h
	\end{array}\right..
\end{eqnarray}

Here
$Q_h=\left(-\frac{\pi}{h},\frac{\pi}{h}\right]^n$ stands for the
$n-$dimensional \textit{Brioullin zone} representation of the $n-$torus $\BR^n/\frac{2\pi}{h}\BZ^n$, as already depicted on Rabin's seminal paper \cite{Rabin82}.
 
With the aid of the \textit{Fourier coefficients}  (cf.~\cite[subsection 5.2.1]{GuerlebeckSproessig97})
\begin{eqnarray}
\label{FourierInversion}
\widehat{\g}_{h,\alpha}(x,t)=\frac{1}{(2\pi)^{\frac{n}{2}}}\int_{Q_h} (\mathcal{F}_{h,\alpha} \g)(\xi,t) e^{-i x \cdot \xi} d\xi
\end{eqnarray}
we are able to derive, in a natural way, the isometric isomorphism $$\mathcal{F}_{\alpha,h}:\ell_2(\BR^n_{h,\alpha};\BC \otimes \cl_{n,n})\rightarrow L_2(Q_h;\BC \otimes \cl_{n,n})$$ with inverse $(\mathcal{F}_{h,\alpha}^{-1} \g)(x,t)=\widehat{\g}_{h,\alpha}(x,t)
$. 

Here and elsewhere $L_2(Q_h;\BC\otimes\cl_{n,n}):=L_2(Q_h)\otimes\left(\BC\otimes \cl_{n,n}\right)$ denotes the $\BC \otimes \cl_{n,n}-$\textit{Hilbert module} endowed by the sesquilinear form
\begin{eqnarray}
\label{BilinearFormQh}\langle \f(\cdot,t),\g(\cdot,t
)\rangle_{Q_h}= \int_{Q_h} \f(\xi,t)^\dag \g(\xi,t) d\xi.
\end{eqnarray}

Moreover, we can mimic the construction provided by \cite[Exercise 3.1.15.]{RuzhanskyT10} \&  \cite[Definition 3.1.25]{RuzhanskyT10} to show that $\mathcal{S}(\BR^n_{h,\alpha};\BC \otimes \cl_{n,n})$ is dense in $\ell_2(\BR^n_{h,\alpha};\BC \otimes \cl_{n,n})$, and that $C^\infty(Q_h;\BC\otimes \cl_{n,n})$ is embedded on $C^\infty(Q_h;\BC \otimes \cl_{n,n})'$, the space of $\BC\otimes \cl_{n,n}-$valued distributions over $Q_h$.

 As a consequence, we uniquely extend the \textit{discrete Fourier transform} (\ref{discreteFh}) as a mapping $\mathcal{F}_{h,\alpha}:\mathcal{S}'(\BR^n_{h,\alpha};\BC \otimes \cl_{n,n})\rightarrow C^\infty(Q_h;\BC \otimes \cl_{n,n})$ by the \textit{Parseval type relation}, involving the sesquilinear forms (\ref{lpInner}) and (\ref{BilinearFormQh}) (cf.~\cite[Definition 3.1.27]{RuzhanskyT10}):
\begin{eqnarray*}
\langle \mathcal{F}_{h,\alpha} \f(\xi,t),\g(\cdot,t
)\rangle_{Q_h}=\left\langle \f(\cdot,t),\widehat{\g}_{h,\alpha}(\cdot,t)\right\rangle_{h,\alpha}, 
\end{eqnarray*}
underlying to $\f(\cdot,t)\in \mathcal{S}'(\BR^n_{h,\alpha};\BC \otimes \cl_{n,n})$ and $\g(\cdot,t) \in C^\infty(Q_h;\BC \otimes\cl_{n,n})$.

With the construction furnished above we can naturally define the convolution between a \textit{discrete distribution} $\f(\cdot,t)$ with membership in $\mathcal{S}'(\BR^n_{h,\alpha};\BC \otimes \cl_{n,n})$, and a \textit{discrete function} $\Phi(x)$ with membership in $\mathcal{S}(\BR^n_{h,\alpha};\BC \otimes \cl_{n,n})$:
\begin{eqnarray}
\label{discreteConvolution} \left(\f(\cdot,t)\star_{h,\alpha} \Phi\right)(x)=\sum_{y\in \BR^n_{h,\alpha}} h^n \Phi(y)\f(y-x,t)
\end{eqnarray}
via the duality condition
\begin{eqnarray*}
\left\langle~ \f(\cdot,t) \star_{h,\alpha} \Phi,\g(\cdot,t)~\right\rangle_{h,\alpha}=\langle~ \f(\cdot,t),\widetilde{\Phi} \star_{h,\alpha} \g(\cdot,t)~\rangle_{h,\alpha}, & \widetilde{\Phi}(x)=[\Phi(-x)]^\dagger,
\end{eqnarray*}
for all $\g(\cdot,t)\in \mathcal{S}(\BR^n_{h,\alpha};\BC\otimes \cl_{n,n})$.

Also, the multiplication of a \textit{continuous distribution} $\mathcal{U}\in C^\infty(Q_h;\BC \otimes\cl_{n,n})'$ by a function $\mathcal{F}_{h,\alpha}\Phi(\xi)$ with membership in $C^\infty(Q_h;\BC \otimes \cl_{n,n})$ can be defined
via the sesquilinear identity
$$ \left\langle~ \left(\mathcal{F}_{h,\alpha}\Phi\right)\mathcal{U} ,\mathcal{F}_{h,\alpha}\g(\cdot,t
)~\right\rangle_{Q_h}= \left\langle~ \mathcal{U} ,\left(\mathcal{F}_{h,\alpha}\Phi\right)^\dag\left(\mathcal{F}_{h,\alpha}\g(\cdot,t
)\right)~\right\rangle_{Q_h}.$$ 

As in \cite[p.~123]{CiaurriGRTV17}, the following \textit{discrete convolution formula} property 
\begin{eqnarray}
\label{ConvolutionFhProperty}\mathcal{F}_{h,\alpha}\left[\f(\cdot,t)\star_{h,\alpha} \Phi\right]=\left(\mathcal{F}_{h,\alpha}\f(\cdot,t)\right)\left(\mathcal{F}_{h,\alpha}\Phi\right)
\end{eqnarray}
that holds at the level of distributions, yields as an immediate consequence of the sequence of identities
\begin{eqnarray*}
\langle~\mathcal{F}_{h,\alpha}\left[\f(\cdot,t)\star_{h,\alpha} \Phi\right],\g(\cdot,t) ~\rangle_{Q_h}&=&
\langle~\f(\cdot,t)\star_{h,\alpha} \Phi, \mathcal{F}_{h,\alpha}^{-1}[\g(\cdot,t)]~ \rangle_{h,\alpha}
\\
&=&\langle~\f(\cdot,t), \widetilde{\Phi}\star_{h,\alpha}\mathcal{F}_{h,\alpha}^{-1}[\g(\cdot,t)]~ \rangle_{h,\alpha} \\
&=&\left\langle~\f(\cdot,t),\mathcal{F}_{h,\alpha}^{-1}\left( \mathcal{F}_{h,\alpha}\widetilde{\Phi}~\g(\cdot,t)\right)~ \right\rangle_{h,\alpha} \\
&=&\left\langle~\mathcal{F}_{h,\alpha}\f(\cdot,t), \mathcal{F}_{h,\alpha}\widetilde{\Phi}~\g(\cdot,t)~ \right\rangle_{h,\alpha} \\
&=&\langle~\left(\mathcal{F}_{h,\alpha}\f(\cdot,t)\right)\left(\mathcal{F}_{h,\alpha}\Phi\right),\g(\cdot,t) ~\rangle_{Q_h}.
\end{eqnarray*}

\subsection{Discrete Dirac-K\"ahler vs. Discrete Laplacian}\label{DiracLaplaceSection}

Let us take now a close look to the Fourier multiplier of $\mathcal{F}_{h,\alpha}\circ (-\Delta_h) \circ \mathcal{F}_{h,\alpha}^{-1}$ encoded by the discrete Laplacian (\ref{discreteLaplacian}).
First, we observe that for $-h \leq \varepsilon\leq h$ the Clifford-valued sesquilinear form $\langle \cdot,\cdot \rangle_{h,\alpha}$ satisfies the
summation property over $\BR_{h,\alpha}^n$ (cf. \cite[p.~536]{FaustinoBayesian17}):
\begin{eqnarray*}
	\sum_{x\in \BR^n_{h,\alpha}}h^n~ \f(x,t)^\dag \g(x +\varepsilon \e_j,t) =\sum_{x\in
		h\BZ^n}h^n~  \f(x-\varepsilon \e_j,t)^\dag \g(x,t).
\end{eqnarray*}

In particular, for the substitutions
\begin{center}
$\f(x,t)\rightarrow e^{-i x \cdot \xi}$ and  $\g(x,t)\rightarrow\Psi(x,t)$
\end{center}
we can conclude that the translation action $x \mapsto \Psi(x+ \varepsilon \e_j,t)$ over $\mathcal{S}(\BR^n_{h,\alpha};\BC \otimes \cl_{n,n})$ gives rise to the property
\begin{eqnarray}
\label{TranslationFh}\mathcal{F}_{h,\alpha} \Psi(\cdot+\varepsilon\e_j,t)=e^{-i\varepsilon \xi_j}\mathcal{F}_{h,\alpha} \Psi(\cdot,t).
\end{eqnarray}

Therefore $\mathcal{F}_{h,\alpha}(\Delta_h \Psi)(\xi,t)=-d_h(\xi)^2\mathcal{F}_{h,\alpha}\Psi(\xi,t)$ (cf. \cite[Subsection 5.2.2]{GuerlebeckSproessig97}), where 
\begin{eqnarray}
\label{SymbolDiscreteLaplace}\displaystyle d_h(\xi)^2=\frac{4}{h^2}\sum_{j=1} \sin^{2}\left(\frac{h\xi_j}{2}\right)
\end{eqnarray}
stands for the Fourier multiplier of $\mathcal{F}_{h,\alpha}\circ (-\Delta_h) \circ \mathcal{F}_{h,\alpha}^{-1}$.

Next, we observe that the sequence of identities  
\begin{eqnarray*}
	\frac{4}{h^2}\sin^{2}\left(\frac{h\xi_j}{2}\right)=&\dfrac{1}{h^2}\left(1-e^{-ih\xi_j}\right)\left(1-e^{ih\xi_j}\right)
	=&\left|\frac{1-e^{-ih\xi_j}}{h}~e^{ih\theta_j}\right|^2
\end{eqnarray*}
hold for every $\xi =(\xi_1,\xi_2,\ldots,\xi_n)$ and $\theta =(\theta_1,\theta_2,\ldots,\theta_n)$ with membership in $Q_h=\left(-\frac{\pi}{h},\frac{\pi}{h}\right]^n$ so that (\ref{SymbolDiscreteLaplace}) may be rewritten as 
$$d_h(\xi)^2=\sum_{j=1}^{n}\left|\frac{1-e^{-ih\xi_j}}{h}~e^{ih\theta_j}\right|^2.$$

In particular, under the choice $\theta_j=(1-\alpha)\xi_j$ the above identity may be expressed in terms of the complex numbers 
\begin{eqnarray*}
	z_{h,\alpha}(\xi_j)=\dfrac{e^{i(1-\alpha) h\xi_j}-e^{-i\alpha h\xi_j}}{h},&\mbox{with}& -\pi<h\xi_j\leq \pi~~~\&~~~0< \alpha< \frac{1}{2}.
\end{eqnarray*}

Moreover, from the set of basic identities 
\begin{eqnarray*}
	|z_{h,\alpha}(\xi_j)|^2&=&\dfrac{1}{2}\left(z_{h,\alpha}(\xi_j)z_{h,\alpha}(\xi_j)^\dagger+z_{h,\alpha}(\xi_j)^\dagger z_{h,\alpha}(\xi_j)\right) \\
	&=& 
	\left(\frac{z_{h,\alpha}(\xi_j)+z_{h,\alpha}(\xi_j)^\dagger}{2}\right)^2-\left(\frac{z_{h,\alpha}(\xi_j)-z_{h,\alpha}(\xi_j)^\dagger}{2}\right)^2 \\
	&=&
	\left(\dfrac{\cos(\alpha h\xi_j)-\cos((1-\alpha) h\xi_j)}{h}\right)^2+\left(\dfrac{\sin(\alpha h\xi_j)+\sin((1-\alpha) h\xi_j)}{h}\right)^2 \\
	&=& 
	\left(\e_{n+j}\dfrac{\cos(\alpha h\xi_j)-\cos((1-\alpha) h\xi_j)}{h}\right)^2+\left(-i\e_j\dfrac{\sin(\alpha h\xi_j)+\sin((1-\alpha) h\xi_j)}{h}\right)^2
\end{eqnarray*} 
one readily has that the Clifford-vector-valued function 
\begin{eqnarray}
\label{DiracSymbol}	
\begin{array}{ccc}
\textbf{z}_{h,\alpha}(\xi)
&=&\displaystyle \sum_{j=1}^n -i\e_j\dfrac{\sin((1-\alpha) h\xi_j)+\sin(\alpha h\xi_j)}{h}+\\
&+&\displaystyle \sum_{j=1}^n \e_{n+j}\dfrac{\cos(\alpha h\xi_j)-\cos((1-\alpha) h\xi_j)}{h}
\end{array}
\end{eqnarray}
satisfies the square condition $\textbf{z}_{h,\alpha}(\xi)^2=d_h(\xi)^2$.

Let us now continue with the class of discrete Dirac-K\"ahler operators $D_\varepsilon$ introduced in (\ref{DiracEqh}).
By means of the $\dagger-$conjugation (\ref{dagconjugation}), one can also define formally the conjugation of $D_\varepsilon$ as follows:
\begin{eqnarray*}
	\label{DiracEqhAdjoint}
	\begin{array}{lll}
D_\varepsilon^\dagger \Psi(x,t)&=&\displaystyle \sum_{j=1}^n -\e_j\frac{\Psi(x+\varepsilon \e_j,t)-\Psi(x-\varepsilon\e_j,t)}{2\varepsilon}+ \\
&+&\displaystyle \sum_{j=1}^n \e_{n+j}\frac{2\Psi(x,t)-\Psi(x+\varepsilon \e_j,t)-\Psi(x-\varepsilon \e_j,t)}{2\varepsilon}.
	\end{array}
\end{eqnarray*}

Here we notice that the combination of the summation property (\ref{TranslationFh}) with the $\dag-$conjugation properties $\e_j^\dag=-\e_j$ and $\e_{n+j}^\dag=\e_{n+j}$  ($j=1,2,\ldots,n$) shows in turn that $D_\varepsilon$ and $D_\varepsilon^\dagger$ 
are self-adjoint with respect to Clifford-valued sesquilinerar form $\langle \cdot,\cdot\rangle_{h,\alpha}$ defined by eq. (\ref{lpInner}), since
\begin{eqnarray*}
\langle D_\varepsilon \f(\cdot,t), \g(\cdot,t) \rangle_{h,\alpha}&=& \langle  \f(\cdot,t), D_\varepsilon \g(\cdot,t) \rangle_{h,\alpha}\\ \langle D_\varepsilon^\dagger \f(\cdot,t), \g(\cdot,t) \rangle_{h,\alpha}&=& \langle  \f(\cdot,t), D_\varepsilon^\dagger \g(\cdot,t) \rangle_{h,\alpha}.
\end{eqnarray*}

As a consequence of the above construction, the Clifford-vector-valued function
$\textbf{z}_{h,\alpha}(\xi)$ defined by (\ref{DiracSymbol}) is the Fourier multiplier of the operator $$\mathcal{F}_{h,\alpha} \circ ((1-\alpha)D_{(1-\alpha)h}-\alpha D_{\alpha h}^\dagger)\circ\mathcal{F}_{h,\alpha}^{-1}.$$

Therefore, for every $0<\alpha<	\frac{1}{2}$ the mapping property $$D_{h,\alpha}:\mathcal{S}(\BR^n_{h,\alpha};\BC \otimes \cl_{n,n})\rightarrow C^\infty(Q_h;\BC \otimes \cl_{n,n})$$
stands for the Dirac-K\"ahler type operator
\begin{eqnarray}
\label{FractionalDiffDirac} D_{h,\alpha}:=(1-\alpha)D_{(1-\alpha)h}-\alpha D_{\alpha h}^\dagger,
\end{eqnarray}
as well as the discrete Laplacian splitting $\left(D_{h,\alpha}\right)^2=-\Delta_h$.

Furthermore, the factorization property
\begin{eqnarray}
\label{FactorizationKleinGordon}  -\Delta_h+m^2=(D_{h,\alpha}-m\gamma)^2
\end{eqnarray}
yields from the set of anti-commuting relations
\begin{eqnarray}
\label{chiralGammaejn}
\gamma \e_j+\e_j \gamma=0 & \gamma \e_{n+j}+ \e_{n+j} \gamma=0 & \gamma^2=+1,
\end{eqnarray}
carrying the Clifford basis elements $\e_j,\e_{n+j}$ ($j=1,2,\ldots,n$), and the pseudo-scalar $\gamma$ defined by eq. (\ref{PseudoScalar}) (cf.~\cite[Proposition 3.1]{FaustinoKGordonDirac16}).

\begin{remark}[Towards a fractional regularization of discrete Dirac operators]	
We would like to stress here that the discretizations $D_\varepsilon$ and $D_{h,\alpha}$, given by eqs. (\ref{DiracEqh}) and (\ref{FractionalDiffDirac}) respectively, are interrelated by the limit formula $$\displaystyle \lim\limits_{\alpha \rightarrow 0} D_{h,\alpha}=D_{h}.$$

On the other hand, the limit property
$$\displaystyle \lim\limits_{\alpha \rightarrow \frac{1}{2}}D_{h,\alpha}=\displaystyle \dfrac{D_{h/2}^+ + D_{h/2}^-}{2}$$
involving the finite difference Dirac operators $D_{h/2}^\pm$ of forward/backward type: 
\begin{eqnarray*}
D_{h/2}^+\Psi(x,t)&=&\sum_{j=1}^n \e_j \dfrac{\Psi\left(x+\frac{h}{2}\e_j,t\right)-\Psi\left(x,t\right)}{h/2} \\
D_{h/2}^-\Psi(x,t)&=&\sum_{j=1}^n \e_j \dfrac{\Psi\left(x,t\right)-\Psi\left(x-\frac{h}{2}\e_j,t\right)}{h/2}.
\end{eqnarray*}
shows us that $D_{h,\alpha}$ may also be seen as a fractional regularization for the discrete Dirac operators on the lattice $\frac{h}{2}\BZ^n$, already considered in the series of papers \cite{Faustino10,FaustinoMonomiality14,FaustinoKGordonDirac16,FaustinoMMAS17}. 

\end{remark}

\begin{remark}[The lattice fermion doubling gap]\label{LatticeDoublingGap}
Since the Fourier multipliers ${\bf z}_{h,\frac{1}{2}}(\xi)$ of $\displaystyle \mathcal{F}_{h,\alpha}\circ \left(\frac{D_{h/2}^+ + D_{h/2}^-}{2}\right)\circ\mathcal{F}_{h,\alpha}^{-1}$ share the same set of zeros of the Fourier multiplier $d_h(\xi)^2$ of $\mathcal{F}_{h,\alpha}\circ \left(-\Delta_h\right)\circ\mathcal{F}_{h,\alpha}^{-1}$ defined in terms of eq. (\ref{SymbolDiscreteLaplace}), we can conclude that the spectrum doubling of $D_{h,\alpha}$ only occurs on the limit $\alpha \rightarrow \frac{1}{2}$ (cf.~\cite[p.~323]{Rabin82}).

At this stage, we have obtained from a multivector perspective that Rabin's homological approach \cite[Section 6]{Rabin82}, based on the geometry of the $n-$torus $\BR^n/\frac{2\pi}{h}\BZ^n$, also works on $\BR_{h,\alpha}^n$ for the discretized Dirac operators $D_{h,\alpha}$.

This is indeed a direct consequence of the so-called \textit{lattice fermion doubling} gap, formulated by Nielsen \& Ninomiya (cf.~\cite{NN81}). We also refer to \cite[subsection 1.1 \& section 4]{FaustinoKGordonDirac16} for further details regarding the discussion of Nielsen--Ninomiya \textit{no-go result}.
\end{remark}

\section{Solution of discretized time-evolution problems}\label{discretizedTimeEvolutions}

\subsection{Discretized Klein-Gordon equations}

In this section we study the solutions of the second-order evolution problems of the type
\begin{eqnarray}
	\label{diffKleinGordon} \left\{\begin{array}{lll} 
		L_t^2\Psi(x,t)= \Delta_h \Psi(x,t)-m^2\Psi(x,t) &, (x,t)\in
		\BR^n_{h,\alpha} \times T
		\\ \ \\
		\Psi(x,0)=\Phi_0(x) &, x\in \BR^n_{h,\alpha}\\ \ \\
		\left[L_t \Psi(x,t)\right]_{t=0}=\Phi_1(x) &, x\in \BR^n_{h,\alpha}
	\end{array}\right.
\end{eqnarray}
on the space-time domain $\BR_{h,\alpha}^n \times T$, from an umbral calculus perspective. 

In terms of the \textit{discrete Fourier transform} (\ref{discreteFh}), the formulation of the time-evolution problem (\ref{diffKleinGordon}) on the momentum space $Q_h \times T$ reads as

\begin{eqnarray}
	\label{diffKleinGordonFourier} \left\{\begin{array}{lll} 
		L_t^2\left[\mathcal{F}_{h,\alpha} \Psi (\xi,t)\right]= -\left(d_h(\xi)^2+m^2\right)\mathcal{F}_{h,\alpha} \Psi (\xi,t) &, (\xi,t)\in Q_h \times T
		\\ \ \\
		\mathcal{F}_{h,\alpha} \Psi (\xi,0)=\mathcal{F}_{h,\alpha} \Phi_0 (\xi) &, \xi\in Q_h\\ \ \\
		\left[L_t\mathcal{F}_{h,\alpha} \Psi (\xi,t)\right]_{t=0}=\mathcal{F}_{h,\alpha} \Phi_1 (\xi)&, \xi\in Q_h
	\end{array}\right..
\end{eqnarray}

With the aid of the functions $(s,t)\mapsto\cosh(tL^{-1}(s))$ and $(s,t)\mapsto \sinh(tL^{-1}(s))$ obtained in {\bf Theorem \ref{EGFcoshsinh}} (see \textsc{Appendix \ref{EGFsection}}) we can describe the solution of the discretized Klein-Gordon equation (\ref{diffKleinGordon}) as a \textit{discrete convolution} on $\BR_{h,\alpha}^n$, endowed by the kernel functions
\begin{eqnarray}
\label{WaveKernels}
\begin{array}{lll}
K_0(x,t)&=&\displaystyle \frac{1}{(2\pi)^{\frac{n}{2}}} \int_{Q_h} \cosh\left(t~L^{-1}\left(i\sqrt{d_h(\xi)^2+m^2}\right)\right) e^{-i x\cdot \xi}~d\xi \\ \ \\
K_1(x,t)&=& \displaystyle \frac{1}{(2\pi)^{\frac{n}{2}}} \int_{Q_h} \frac{\sinh\left(t~L^{-1}\left(i\sqrt{d_h(\xi)^2+m^2}\right)\right)}{i\sqrt{d_h(\xi)^2+m^2}} e^{-i x\cdot \xi}~d\xi.
\end{array}
\end{eqnarray}

More precisely, for the discretized wave propagators defined in terms of the \textit{discrete convolution formulae}
\begin{eqnarray}
\label{WavePropagators}
\begin{array}{lll}
\cosh\left(t~L^{-1}\left(\sqrt{\Delta_h-m^2}\right)\right)\Phi(x)&=&\displaystyle \sum_{y \in \BR^n_{h,\alpha}} h^n \Phi(y)K_0(x-y,t)
\\
\displaystyle \frac{\sinh\left(t~L^{-1}\left(\sqrt{\Delta_h-m^2}\right)\right)}{\sqrt{\Delta_h-m^2}}\Phi(x)&=&\displaystyle \sum_{y \in \BR^n_{h,\alpha}} h^n\Phi(x) K_1(x-y,t)
\end{array}
\end{eqnarray}
we are able to mimic the so-called wave Duhamel formula \textit{in continuum} (cf.~\cite[\textsc{Exercise 2.22}]{Tao06} \& \cite[p. 71]{Tao06}). That corresponds to the following theorem:

\begin{theorem}\label{mainResultdiffKleinGordon}
Let $\Phi_0$ and $\Phi_1$ be two Clifford-valued functions membership in $\mathcal{S}(\BR_{h,\alpha}^n;\BC \otimes \cl_{n,n})$, and 
	$K_0$, $K_1$ be the kernel functions defined by the integral formulae (\ref{WaveKernels}).
	 Then we have the following:
	\begin{enumerate}
		\item[{\bf (i)}] The function
		 \begin{eqnarray}
		\label{ansatzFhKleinGordoncoshsinh}
		\begin{array}{ccc}
		\mathcal{F}_{h,\alpha} \Psi (\xi,t)&=&\cosh\left(t L^{-1}\left(i\sqrt{d_h(\xi)^2+m^2}\right)\right)	\mathcal{F}_{h,\alpha} \Phi_0 (\xi)+ \\ \ \\
		&+&\displaystyle  \dfrac{\sinh\left(t L^{-1}\left(i\sqrt{d_h(\xi)^2+m^2}\right)\right)}{i\sqrt{d_h(\xi)^2+m^2}}	\mathcal{F}_{h,\alpha} \Phi_1 (\xi)
		\end{array}
		\end{eqnarray}
		solves the time-evolution problem (\ref{diffKleinGordonFourier}).
		\item[{\bf (ii)}] The ansatz \begin{eqnarray}
		\label{ansatzKleinGordoncoshsinh}
		\begin{array}{ccc}
		\Psi (x,t)&=&\cosh\left(t L^{-1}\left(\sqrt{\Delta_h-m^2}\right)\right) \Phi_0 (x)+ \\ \ \\
		&+&\displaystyle  \dfrac{\sinh\left(t L^{-1}\left(\sqrt{\Delta_h-m^2}\right)\right)}{\sqrt{\Delta_h-m^2}} \Phi_1 (x)
		\end{array}
		\end{eqnarray}
		solves the discretized Klein-Gordon equation (\ref{diffKleinGordon}). 
	\end{enumerate} 
\end{theorem}

\proof

\vspace{0.20cm}

\textbf{Proof of (i):}

In the shed of \textbf{Theorem \ref{EGFhypercomplex}} let us now take a close look to the ansatz functions of the type
\begin{eqnarray}
\label{ansatzFh}
\begin{array}{ccc}
\mathcal{F}_{h,\alpha} \Psi (\xi,t)&=&\G(e^{i\phi}r\omega,t)\mathcal{F}_{h,\alpha} \Phi_+  (\xi)+ \G(-e^{i\phi}r\omega,t)\mathcal{F}_{h,\alpha} \Phi_- (\xi).
\end{array}
\end{eqnarray}

From the eigenvalue property (\ref{EGFLtEigenvalue}) one readily obtains 
 that (\ref{ansatzFh}) satisfies the equation $$L_t^2\left[\mathcal{F}_{h,\alpha} \Psi (\xi,t)\right]=-\left(d_h(\xi)^2+m^2\right)\mathcal{F}_{h,\alpha} \Psi (\xi,t)$$
 whenever $\phi=\frac{\pi}{2}$, $r=\sqrt{d_h(\xi)^2+m^2}$ and $\omega^2=+1$. 
  
It is also straightforward to see that the initial conditions of the evolution problem (\ref{diffKleinGordonFourier}) lead to the system of equations
\begin{eqnarray*}
	\mathcal{F}_{h,\alpha} \Phi_0 (\xi)&=&\mathcal{F}_{h,\alpha} \Phi_+ (\xi)+\mathcal{F}_{h,\alpha} \Phi_- (\xi) \\ 	\mathcal{F}_{h,\alpha} \Phi_1 (\xi)&=&i\omega \sqrt{d_h(\xi)^2+m^2}~~\mathcal{F}_{h,\alpha} \Phi_+ (\xi)-i\omega\sqrt{d_h(\xi)^2+m^2}~~\mathcal{F}_{h,\alpha} \Phi_- (\xi).
\end{eqnarray*}

Solving the above system of equations in order to $\mathcal{F}_{h,\alpha} \Phi_\pm (\xi)$, it readily follows that
\begin{eqnarray*}
	\mathcal{F}_{h,\alpha} \Phi_\pm (\xi)&=&\dfrac{1}{2}\mathcal{F}_{h,\alpha} \Phi_0 (\xi)\pm\dfrac{\omega}{2i\sqrt{d_h(\xi)^2+m^2}}\mathcal{F}_{h,\alpha} \Phi_1 (\xi).  
\end{eqnarray*}

Therefore, we can recast (\ref{ansatzFh}) as
\begin{eqnarray}
\label{ansatzFhKleinGordon}
\begin{array}{ccc}
	\mathcal{F}_{h,\alpha} \Psi (\xi,t)&=&\dfrac{\G\left(i\omega\sqrt{d_h(\xi)^2+m^2},t\right)+\G\left(-i\omega\sqrt{d_h(\xi)^2+m^2},t\right)}{2}	\mathcal{F}_{h,\alpha} \Phi_0 (\xi)+\nonumber \\ \ \\
&+&\displaystyle \omega \dfrac{\G\left(i\omega\sqrt{d_h(\xi)^2+m^2},t\right)-\G\left(-i\omega\sqrt{d_h(\xi)^2+m^2},t\right)}{2i\sqrt{d_h(\xi)^2+m^2}}	\mathcal{F}_{h,\alpha} \Phi_1 (\xi).
\end{array}
\end{eqnarray}

Moreover, from \textbf{Theorem \ref{EGFhypercomplex}} we immediately get that eq. (\ref{ansatzFhKleinGordon}) is equivalent to eq.~(\ref{ansatzFhKleinGordoncoshsinh}).

\vspace{0.20cm}
\textbf{Proof of (ii):} 

By applying the \textit{discrete Fourier transform} $\mathcal{F}_{h,\alpha}$ to both sides of (\ref{WavePropagators}) the {\bf Proof of (ii)} follows straightforwardly from the \textit{discrete convolution property} (\ref{ConvolutionFhProperty}),
and from the fact that the function $\Psi(x,t)$ defined by eq. (\ref{ansatzKleinGordoncoshsinh}) is also a solution of the equation
$$\displaystyle L_t^2\left[\mathcal{F}_{h,\alpha} \Psi (\xi,t)\right]=-\left(d_h(\xi)^2+m^2\right)\mathcal{F}_{h,\alpha} \Psi (\xi,t).$$
\qed

\subsection{Discretized Dirac equations}

Let us now look to discretized version of the Dirac equation
\begin{eqnarray}
\label{diffDirac} \left\{\begin{array}{lll} 
-iL_t \Psi (x,t)= \left(D_{h,\alpha}-m\gamma\right) \Psi (x,t) &, (x,t)\in \BR_{h,\alpha}^n \times T 
\\ \ \\
\Psi (x,0)=\Phi_0 (x) &, x\in \BR_{h,\alpha}^n
\end{array}\right.,
\end{eqnarray}
carrying the discrete Dirac operator $D_{h,\alpha}$ introduced in (\ref{FractionalDiffDirac}).

From the framework developed on the previous sections and on \textsc{Appendix \ref{EGFsection}}, the solution of (\ref{diffDirac}) can be easily found. In concrete, the formal solution of (\ref{diffDirac}) provided by the operational formula
$\displaystyle \Psi(x,t)={\bf G}\left(iD_{h,\alpha}-im\gamma,t\right)\Phi_0(x)$
is a direct consequence of the set of identities
\begin{eqnarray*}
	-iL_t\left[\mathcal{F}_{h,\alpha} \Psi (\xi,t)\right]&=& \left(\textbf{z}_{h,\alpha}(\xi)-m\gamma\right)\mathcal{F}_{h,\alpha} \Psi (\xi,t) \\
	{\bf G}\left(i\textbf{z}_{h,\alpha}(\xi)-im\gamma,0\right)&=&1 \\
	L_t{\bf G}\left(i\textbf{z}_{h,\alpha}(\xi)-im\gamma,t\right)&=&i\left(\textbf{z}_{h,\alpha}(\xi)-m\gamma\right){\bf G}\left(i\textbf{z}_{h,\alpha}(\xi)-im\gamma,t\right)
\end{eqnarray*}
associated to the Fourier multiplier of $\textbf{z}_{h,\alpha}(\xi)$ of  $\mathcal{F}_{h,\alpha}\circ D_{h,\alpha}\circ\mathcal{F}_{h,\alpha}^{-1}$ (see equation (\ref{DiracSymbol})) and the EGF ${\bf G}(\s,t)$ (see equation (\ref{EGFLt})). Then, the following corollary is rather obvious:

\begin{corollary}\label{mainResultdiffDirac}
Let $\Phi_0$ be a function with membership in $\mathcal{S}(\BR_{h,\alpha}^n;\BC \otimes \cl_{n,n})$. Then, under the first order condition $$[L_t\Psi(x,t)]_{t=0}=i\left(D_{h,\alpha}-m\gamma\right)\Phi_0(x)$$ 
the ansatz (\ref{ansatzKleinGordoncoshsinh}) solves the discretized Dirac equation (\ref{diffDirac}).
\end{corollary}

\proof
First, we recall that the square relation $$(\textbf{z}_{h,\alpha}(\xi)-m\gamma)^2=d_h(\xi)^2+m^2$$ involving the Fourier multipliers $d_h(\xi)^2$ and $\textbf{z}_{h,\alpha}(\xi)$, defined by equations (\ref{SymbolDiscreteLaplace}) resp. (\ref{DiracSymbol}),  yields as a direct consequence of the factorization property (\ref{chiralGammaejn}) involving the Clifford generators $\e_j,\e_{n+j}$ ($j=1,2,\ldots,n$), and the pseudo-scalar $\gamma$ defined by eq. (\ref{PseudoScalar}).
Then $$\displaystyle \omega=\dfrac{i\textbf{z}_{h,\alpha}(\xi)-im\gamma}{i\sqrt{d_h(\xi)^2+m^2}}$$ is a unitary vector of $\BC\otimes\cl_{n,n}$ satisfying the property $\omega^2=+1$.
Thus, from  \textbf{Theorem \ref{EGFhypercomplex}}
\begin{eqnarray*}
{\bf G}\left(i\textbf{z}_{h,\alpha}(\xi)-im\gamma,t\right)&=&\cosh\left(tL^{-1}\left(i\sqrt{d_h(\xi)^2+m^2}\right)\right)+\\
&+&\dfrac{\sinh\left(tL^{-1}\left(i\sqrt{d_h(\xi)^2+m^2}\right)\right)}{i\sqrt{d_h(\xi)^2+m^2}}\left(i\textbf{z}_{h,\alpha}(\xi)-im\gamma\right).
\end{eqnarray*}

Moreover, from the property
$$\displaystyle \mathcal{F}_{h,\alpha}[i(D_{h,\alpha}-m\gamma)\Phi_0](\xi)=\left(i\textbf{z}_{h,\alpha}(\xi)-im\gamma\right)\mathcal{F}_{h,\alpha}\Phi_0(\xi)$$ it readily follows from direct application of statement {\bf (ii)} of {\bf Theorem \ref{mainResultdiffKleinGordon}} that ${\bf G}\left(iD_{h,\alpha}-im\gamma,t\right)\Phi_0(x)$ -- a formal solution of the discretized Dirac equation (\ref{diffDirac}) -- equals to the ansatz (\ref{ansatzFhKleinGordoncoshsinh}), whenever
$\displaystyle \Phi_1(x)=i(D_{h,\alpha}-m\gamma)\Phi_0(x).$
\qed

\begin{remark}[The Zassenhaus formula gap]
	The framework that we have considered here to describe formally the solutions of the discretized Dirac equation (\ref{diffDirac}) may be seen as a multivector extension of the framework obtained in terms of Pauli matrices by Datolli and his collaborators on the papers \cite{DattoliSGHP15,DattolliT15}.
		The major difference here lies in fact that we have considered the EGF provided by {\bf Theorem \ref{EGFhypercomplex}} to rid the limitations associated to the operational representation of Dirac type propagators by means of the Zassenhaus formula (cf.~\cite[p.~8]{DattoliSGHP15}). 
\end{remark}

\section{Further Applications}\label{FurtherApplications}

\subsection{A space-time Fourier inversion formula based on Chebyshev polynomials}\label{HypergeometricChebyshev}

Let us now discuss a difference-difference version of the evolution problems (\ref{diffKleinGordon}) and (\ref{diffDirac}) on the lattice $\BR_{h,\alpha}^n\times \frac{\tau}{2}\BZ_{\geq 0}$, associated to the finite diference operator (\ref{DifferenceDifferenceLt}) defined on subsection \ref{outlinePaper}.
From direct application of \textbf{Theorem \ref{mainResultdiffKleinGordon}} it can be easily seen that
\begin{eqnarray*}
	\mathcal{F}_{h,\alpha}\Psi(\xi,t)&=&\cos\left(\dfrac{2t}{\tau}\sin^{-1}\left(\frac{\tau}{2}\sqrt{d_h(\xi)^2+m^2}\right)\right)\mathcal{F}_{h,\alpha}\Phi_0(\xi)+\\
	&+&\frac{\sin\left(\dfrac{2t}{\tau}\sin^{-1}\left(\dfrac{\tau}{2}\sqrt{d_h(\xi)^2+m^2}\right)\right)}{\sqrt{d_h(\xi)^2+m^2}}\mathcal{F}_{h,\alpha}\Phi_1(\xi)
\end{eqnarray*}
solves the second-order time-evolution problem (\ref{diffKleinGordonFourier}) on the momentum space $Q_h\times \BZ_{\geq 0}$ (see statement {\bf (i)}).

Here we recall that from the inverse trigonometric relation $\sin^{-1}(z)=\cos^{-1}(\sqrt{1-z^2})$, that holds for every $0\leq z\leq 1$, we can recognize that the above identity may be expressed in terms of Chebyshev polynomials of first and second kind (cf.~\cite[p.~170]{MasonChebyshev93}) for values of $\tau$ satisfying the following condition:
$$ 0\leq \sqrt{d_h(\xi)^2+m^2} \leq \frac{2}{\tau}.$$

That is,
\begin{eqnarray}
\label{ChebyshevExpansion}
\begin{array}{lll}
\mathcal{F}_{h,\alpha}\Psi(\xi,t)&=&
T_{\frac{2t}{\tau}}\left(\sqrt{1-\dfrac{\tau^2}{4}\left(d_h(\xi)^2+m^2\right)}\right)\mathcal{F}_{h,\alpha}\Phi_0(\xi)+\\ \ \\
&+&\dfrac{\tau}{2}U_{\frac{2t}{\tau}-1}\left(\sqrt{1-\dfrac{\tau^2}{4}\left(d_h(\xi)^2+m^2\right)}\right)\mathcal{F}_{h,\alpha}\Phi_1(\xi),
\end{array}
\end{eqnarray}
with
$\displaystyle T_{k}(\lambda)=\cos(k\cos^{-1}(\lambda))$ resp. $\displaystyle U_{k-1}(\lambda)=\frac{\sin(k\cos^{-1}(\lambda))}{\sqrt{1-\lambda^2}}$.

We recall here that the Chebyshev polynomials of first and second kind, $T_k$ resp. $U_{k-1}$, admit the following Cauchy principal value representations (cf.~\cite[subsection 4.1.]{MasonChebyshev93}):
\begin{eqnarray}
\label{ChebyshevIntegrals}
\begin{array}{lll}
\displaystyle \int_{-1}^{1} \frac{U_{k-1}(s)}{s-\lambda}(1-s^2)^{\frac{1}{2}} ds&=&-\pi T_{k}(\lambda)\\ \ \\
\displaystyle \int_{-1}^{1} \frac{T_{k}(s)}{s-\lambda}(1-s^2)^{-\frac{1}{2}} ds&=&\pi U_{k-1}(\lambda).
\end{array}
\end{eqnarray}

In particular, for the change of variable $s=\cos\left(\frac{\omega\tau}{2}\right)$ ($0\leq \omega\leq \frac{2\pi}{\tau}$), the sequence of identities
\begin{eqnarray}
\label{ChebyshevIntegralsTrigonometric}
\begin{array}{lll}
T_{\frac{2t}{\tau}}(\lambda)&=&\displaystyle- \frac{\tau}{2\pi} \int_{0}^{\frac{2\pi}{\tau}} \frac{\sin\left(\frac{\omega\tau}{2}\right)}{\cos\left(\frac{\omega\tau}{2}\right)-\lambda} ~\sin(\omega t)d\omega \\ \ \\
&=&\displaystyle \frac{\tau}{4\pi} \int_{-\frac{2\pi}{\tau}}^{\frac{2\pi}{\tau}} \frac{-i\sin\left(\frac{\omega\tau}{2}\right)}{\cos\left(\frac{\omega\tau}{2}\right)-\lambda}~e^{-i\omega t} d\omega \\ \ \\
U_{\frac{2t}{\tau}-1}(\lambda)&=&\displaystyle \frac{\tau}{2\pi} \int_{0}^{\frac{2\pi}{\tau}} \frac{1}{\cos\left(\frac{\omega\tau}{2}\right)-\lambda}~\cos(\omega t) d\omega \\ \ \\
&=&\displaystyle \frac{\tau}{4\pi} \int_{-\frac{2\pi}{\tau}}^{\frac{2\pi}{\tau}} \frac{1}{\cos\left(\frac{\omega\tau}{2}\right)-\lambda} ~e^{-i
	\omega t}d\omega.	
\end{array}
\end{eqnarray}
follow straightforwardly from parity arguments carrying the conjugation of the complex exponential function $e^{i\omega t}=\cos(\omega t)+i\sin(\omega t)$ (cf.~\cite[p.~173]{MasonChebyshev93}).

Thus, based on (\ref{ChebyshevIntegralsTrigonometric}) one finds that (\ref{ChebyshevExpansion}) admits the integral representation formula
\begin{eqnarray}
\label{ChebyshevHypersingular}
\mathcal{F}_{h,\alpha}\Psi(\xi,t)&=&\frac{\tau}{4\pi} \int_{-\frac{2\pi}{\tau}}^{\frac{2\pi}{\tau}} \frac{-i\sin\left(\frac{\omega\tau}{2}\right)\mathcal{F}_{h,\alpha}\Phi_0(\xi)+\frac{\tau}{2}\mathcal{F}_{h,\alpha}\Phi_1(\xi)}{\cos\left(\frac{\omega\tau}{2}\right)-\sqrt{1-\frac{\tau^2}{4}\left(d_h(\xi)^2+m^2\right)}} ~e^{-i
	\omega t}d\omega.
\end{eqnarray}

Thereby, in the view of the Fourier inversion formula for $\mathcal{F}_{h,\alpha}$ provived by (\ref{FourierInversion}), the \textit{discrete convolution formula} provided by statement {\bf (iii)} of 
\textbf{Theorem \ref{mainResultdiffKleinGordon}} may be reformulated as a \textit{space-time Fourier inversion formula} over $Q_h \times \left(-\frac{2\pi}{\tau},\frac{2\pi}{\tau}\right]$:

\begin{eqnarray*}
\label{ChebyshevHypersingularMomentum}
\Psi(x,t)=\frac{\tau}{(2\pi)^{\frac{n}{2}+2}} \int_{Q_h}\int_{-\frac{2\pi}{\tau}}^{\frac{2\pi}{\tau}} \frac{-i\sin\left(\frac{\omega\tau}{2}\right)\mathcal{F}_{h,\alpha}\Phi_0(\xi)+\frac{\tau}{2}\mathcal{F}_{h,\alpha}\Phi_1(\xi)}{\cos\left(\frac{\omega\tau}{2}\right)-\sqrt{1-\frac{\tau^2}{4}\left(d_h(\xi)^2+m^2\right)}} ~e^{-i
	(\omega t+x\cdot \xi)}d\omega d\xi.
\end{eqnarray*}

Moreover, through
the substitution $\mathcal{F}_{h,\alpha}\Phi_1(\xi)=i({\bf z}_{h,\alpha}(\xi)-m\gamma)\mathcal{F}_{h,\alpha}\Phi_0(\xi)$ on the right-hand side of the above equality, we recognize that the above integral representation over $Q_h \times \left(-\frac{2\pi}{\tau},\frac{2\pi}{\tau}\right]$ also fulfils for the discretized Dirac equation (\ref{diffDirac}) (see statement {\bf (ii)} of \textbf{Corollary \ref{mainResultdiffDirac}}).

\begin{remark}[Connection with the discrete Cauchy-Kovaleskaya extension]\label{discreteCKremark}
The solution of the discretized Dirac equation that we have considered here for the finite difference operator $L_t$ defined by eq. (\ref{DifferenceDifferenceLt}) resembles the construction considered by Constales and De Ridder on the paper \cite{ConstalesDeRidder14}, from a discrete Fourier analysis perspective.
 Here we recall that from the isomorphism (cf.~\cite{Rabin82}) $$Q_h \times \left(-\frac{2\pi}{\tau},\frac{2\pi}{\tau}\right] \cong \left(\BR^{n} /\frac{2\pi}{h}\BZ^n\right) \times \left(\BR /\frac{4\pi}{\tau}\BZ\right),$$ the resulting integral representation formula may be interpreted as a \textit{space-time toroidal Fourier transform} (cf.~\cite[section 3 of Part II]{RuzhanskyT10}).
\end{remark}

\subsection{Connection with the discrete heat semigroup}\label{discreteHeatsemigroup}

In this subsection we will explore the connection between the solutions of the discretized Klein-Gordon and Dirac equations obtained in subsection \ref{HypergeometricChebyshev}, and the
solution of the differential-difference heat equation
\begin{eqnarray}
	\label{diffHeat} \left\{\begin{array}{lll} 
		\partial_s \Psi (x,s)= \Delta_h \Psi (x,s) &, (x,s)\in 
		\BR^n_{h,\alpha} \times [0,\infty) 
		\\ \ \\
		\Psi (x,0)=\Phi (x) &, x\in \BR_{h,\alpha}^n
	\end{array}\right.
\end{eqnarray}
by means of the \textit{discrete heat semigroup} 
$\{\exp(s\Delta_h)\}_{s\geq 0}$.
Before we proceed, we will revisit the construction of the \textit{discrete heat kernel} obtained by Baaske et al in \cite{BaaskeBRS14}.
Along the same lines as in \cite[section 2.]{CiaurriGRTV17} one can show that $\exp(s\Delta_h)$ may be expressed in terms of the \textit{discrete 
convolution formula}
\begin{eqnarray}
\label{discreteConvolutionHeat}\exp(s\Delta_h)\Phi(x)&=&\sum_{y \in \BR_{h,\alpha}^n}  h^n\Phi(y)K(x-y,s),
\end{eqnarray}
involving the kernel function
\begin{eqnarray}
\label{IntegralDiscreteHeat}K(x,s)=\frac{1}{(2\pi)^{\frac{n}{2}}} \int_{Q_h} e^{-\frac{4s}{h^2}d_h(\xi)^2} e^{i x \cdot \xi} d\xi.
\end{eqnarray}

On the other hand, in view of integral representation formula for the \textit{modified Bessel functions of the first kind} $I_k(u)$:
$$
I_k(u)=\frac{1}{\pi}\int_{0}^\pi e^{u\cos(\theta)}\cos(k\theta)d\theta=\frac{1}{2\pi}\int_{-\pi}^\pi e^{u\cos(\theta)}e^{-ik\theta}d\theta,
$$
and the identity associated to the Fourier multipliers (\ref{SymbolDiscreteLaplace}):
$$ 
d_h(\xi)^2=\sum_{j=1}^n \frac{2}{h^2}\left(1-\cos(h\xi_j)\right)
$$
we thereby obtain the closed formula 
\begin{eqnarray}
\label{ClosedFormulaBessel} K(x,s)&=&\frac{(2\pi)^{\frac{n}{2}}}{h^{n}}e^{-\frac{2ns}{h^2}}  I_{\frac{x_1}{h}}\left(\frac{2s}{h^2}\right)I_{\frac{x_2}{h}}\left(\frac{2s}{h^2}\right) \ldots I_{\frac{x_n}{h}}\left(\frac{2s}{h^2}\right),
\end{eqnarray}
after the change of variables $\xi_j=\frac{\theta_j}{h}$ ($-\pi <\theta_j\leq \pi$) on (\ref{IntegralDiscreteHeat}).

Next, let us turn again our attention to the \textit{space-time Fourier inversion formula} (\ref{ChebyshevHypersingular}) derived on {\bf section \ref{HypergeometricChebyshev}}.
Starting from the Laplace transform identity (cf.~\cite[p.~21]{SamkoEtAl93} \& \cite[p.~282]{SaxenaEtAl02}) 
\begin{eqnarray}
\label{LaplaceIdentityMittagLeffler}
\int_{0}^\infty e^{p\lambda^2} p^{\beta-1}E_{\alpha,\beta}\left(s p^{\alpha}\right)dp= \dfrac{\lambda^{-2\beta}}{1-s \lambda^{-2\alpha}},~~ \Re(\lambda^2)>|s|^{\frac{1}{\alpha}}~\&~\Re(\beta)>0
\end{eqnarray}
 involving the \textit{generalized Mittag-Leffler functions} (cf.~\cite[subsection 4.2.]{FaustinoBayesian17})
\begin{eqnarray*} \label{MittagLeffler}E_{\alpha,\beta}(z)=\sum_{k=0}^{\infty}\frac{z^k}{\Gamma(\beta+\alpha k)}, &\mbox{for}& \Re(\alpha)>0~\&~\Re(\beta)>0,
\end{eqnarray*}
we realize that 
\begin{eqnarray*}
\frac{1}{\cos\left(\frac{\omega\tau}{2}\right)-\sqrt{1-\frac{\tau^2}{4}\left(d_h(\xi)^2+m^2\right)}}=\\ 
=-\int_{0}^\infty e^{-\frac{p\tau^2}{4}d_h(\xi)^2}~\frac{E_{\frac{1}{2},\frac{1}{2}}\left(\cos\left(\frac{\omega\tau}{2}\right) \sqrt{p}\right)}{\sqrt{p}}~e^{p\left(1-\frac{\tau^2}{4}m^2\right)}dp
\end{eqnarray*}
so that
 (\ref{ChebyshevHypersingular}) is equivalent to
\begin{eqnarray*}
\mathcal{F}_{h,\alpha}\Psi(\xi,t)=-\frac{\tau}{(2\pi)^{\frac{n}{2}+2}}  \int_{-\frac{2\pi}{\tau}}^{\frac{2\pi}{\tau}} \int_{0}^\infty e^{-\frac{p\tau^2}{4}d_h(\xi)^2}~
\frac{E_{\frac{1}{2},\frac{1}{2}}\left(\cos\left(\frac{\omega\tau}{2}\right) \sqrt{p}\right)}{\sqrt{p}} \times \\ \ \\ \times \left(-i\sin\left(\frac{\omega\tau}{2}\right)\mathcal{F}_{h,\alpha}\Phi_0(\xi)+\frac{\tau}{2}\mathcal{F}_{h,\alpha}\Phi_1(\xi)\right)~e^{p\left(1-\frac{\tau^2}{4}m^2\right)}~e^{-i
	\omega t}dp d\omega. 
\end{eqnarray*}

By making again use of the inversion formula (\ref{FourierInversion}) associated to  $\mathcal{F}_{h,\alpha}$ and after some straightforward simplifications involving the interchanging on the order of integration, there holds 
\begin{eqnarray}
\label{LaplaceSolution}
\Psi(x,t)= 
 \int_{0}^{\infty} \exp\left({\frac{p\tau^2}{4}\Delta_h}\right) [\Phi(x,t;p)]dp,
\end{eqnarray}
with
\begin{eqnarray*}
\Phi(x,t;p)&=&-\frac{\tau}{4\pi} \int_{-\frac{2\pi}{\tau}}^{\frac{2\pi}{\tau}} \left({-i\sin\left(\frac{\omega\tau}{2}\right)\Phi_0(x)+\frac{\tau}{2}\Phi_1(x)}\right)\times \\ \ \\ &\times& \dfrac{E_{\frac{1}{2},\frac{1}{2}}\left(\cos\left(\frac{\omega\tau}{2}\right) \sqrt{p}\right)}{\sqrt{p}}~e^{p\left(1-\frac{\tau^2}{4}m^2\right)}~e^{-i
	\omega t}d\omega.
\end{eqnarray*}

Here one notice that the substitution $s=\frac{u\tau^2}{4}$ on (\ref{diffHeat}) 
reveals that the action $\exp\left({\frac{p\tau^2}{4}\Delta_h}\right) [\Phi(x,t;p)]$ corresponds to the \textit{discrete convolution formula} (\ref{discreteConvolutionHeat}), written in terms of the \textit{discrete heat kernel} $K\left( x,\frac{p\tau^2}{4}\right)$. The later may be computed from (\ref{ClosedFormulaBessel}) as a product of the \textit{modified Bessel functions of the first kind} $I_k(u)$.
 Thus, as in \cite{CiaurriGRTV17} the solution of the discretized Klein-Gordon equation on the lattice $\BR_{h,\alpha}^n \times \frac{\tau}{2}\BZ_{\geq 0}$ may be recovered from the \textit{discrete heat semigroup}.

On the other hand, from the set of identities (cf.~\cite[p.~281]{SaxenaEtAl02})
\begin{eqnarray*}
E_{\alpha,\beta}(u)&=&\frac{1}{u}E_{\alpha,\beta-\alpha}(u)-\frac{1}{u}\frac{1}{\Gamma(\beta-\alpha)} \\
E_{\frac{1}{2},1}(u)&=&e^{u^2}\mbox{erfc}(-u)
\end{eqnarray*}
involving the \textit{generalized Mittag-Leffler functions} and the \textit{complementary error function}
$$
\mbox{erfc}(-u)=\frac{2}{\sqrt{\pi}}\int_{-u}^{\infty} e^{-q^2}dq
$$
we realize, after a short computation by means of the trigonometric identities ($t\in \tau\mathbb{Z}_{\geq 0}$ \& $-\frac{2\pi}{\tau}<\omega\leq \frac{2\pi}{\tau}$):
\begin{eqnarray*}
1+\cos^2\left(\frac{\omega\tau}{2}\right)=2\cos(\omega \tau),& \displaystyle \sin\left(\frac{\omega\tau}{2}\right)\cos\left(\frac{\omega\tau}{2}\right)=\dfrac{1}{2}\sin(\omega \tau)~~\& & \sin\left(\dfrac{2\pi t}{\tau}\right)=0,
\end{eqnarray*}
and on parity arguments that the function $\Phi(x,t;p)$, defined as above, equals to
\begin{eqnarray*}
\Phi(x,t;p)=-\frac{\tau}{8\pi} \int_{-\frac{2\pi}{\tau}}^{\frac{2\pi}{\tau}} \left({-i\sin(\omega \tau)\Phi_0(x)+\tau \cos\left(\frac{\omega\tau}{2}\right)\Phi_1(x)}\right)\times \\ \times~ \mbox{erfc}\left(~-\cos\left(\frac{\omega\tau}{2}\right) \sqrt{p}~\right){e^{p\left(2\cos(\omega \tau)-\frac{\tau^2}{4}m^2\right)}}~e^{-i
	\omega t}d\omega.
\end{eqnarray*}

\begin{remark}\label{LaplaceWeierstrassRemark}
One notice here that the Laplace operational identity (\ref{LaplaceSolution}) slightly differs from the one considered in \cite[section 2]{CiaurriGRTV17} for the operational representation of an analogue for the Poisson type semigroup.
 Here we have considered a \textit{time Fourier inversion formula} over $\left(-\frac{2\pi}{\tau},\frac{2\pi}{\tau}\right]$, that results from the identity 
 \begin{eqnarray*}
 	\frac{1}{\sqrt{p}}
\left(E_{\frac{1}{2},\frac{1}{2}}\left(\cos\left(\frac{\omega\tau}{2}\right)\sqrt{p}\right)+\frac{1}{\sqrt{\pi}}\right)= \cos\left(\frac{\omega\tau}{2}\right)~e^{p\cos^2\left(\frac{\omega\tau}{2}\right)}~\mbox{erfc}\left(~-\cos\left(\frac{\omega\tau}{2}\right) \sqrt{p}~\right)
 \end{eqnarray*}
involving the \textit{complementary error function} $\mbox{erfc}(-u)$,
  instead of the subordination formula
 $$e^{-\beta t}=\frac{t}{2\sqrt{\pi}}\int_0^{\infty} \frac{e^{-\frac{t^2}{4v}}}{v^{\frac{3}{2}}}e^{-v\beta^2}dv~~~(\beta>0)
 $$
 endowed by the kernel of the Weierstra\ss ~transform.
\end{remark}

\subsection{A discrete fractional calculus insight}\label{DiscreteFractionalCalculus}

In section \ref{discretizedTimeEvolutions} we have shown that a simple operational substitution $s \rightarrow \sqrt{\Delta_h-m^2}$ on the functions 
\begin{center}
$\displaystyle \cosh(tL^{-1}(s))$ and $\displaystyle \dfrac{\sinh(tL^{-1}(s))}{s}$
\end{center}
allows us to express, in a simple way, the solutions of the discretized Klein-Gordon and Dirac equations, (\ref{diffKleinGordon}) resp. (\ref{diffDirac}).
Such characterization may be reformulated in terms of the fractional operators
\begin{eqnarray}
\label{FractionalKleinGordon}
\begin{array}{lll}
\displaystyle \left(-\Delta_h+m^2\right)^{-\alpha}\Phi(y)&=&\displaystyle \frac{1}{\Gamma(\alpha)}\int_0^{\infty} e^{-tm^2}\exp\left(t\Delta_h\right)\left[\Phi(y)\right]~t^\alpha \frac{dt}{t}
\end{array}
\end{eqnarray} 
for values $0<\alpha<\frac{1}{2}$, in spite of the right-hand side of (\ref{FractionalDiffDirac}) does not absolutely converges for $\alpha=\frac{1}{2}$ in the \textit{massless limit} $m\rightarrow 0$. This is due to the fact that the Fourier multiplier $\left(d_h(\xi)^2\right)^{-\frac{1}{2}}$ of $\mathcal{F}_{h,\alpha}\circ (-\Delta_{h})^{-\frac{1}{2}} \circ \mathcal{F}_{h,\alpha}^{-1}$ does not belong to the space $C^{\infty}(Q_h;\BC\otimes\cl_{n,n})$. 

The following result, which complements the construction of {\bf Theorem \ref{mainResultdiffKleinGordon}},
provides us an alternative way to obtain a solution for the discretized Klein-Gordon equation (\ref{diffKleinGordon}) as a \textit{discrete convolution formula} endowed by the kernel functions
\begin{eqnarray}
\label{WaveKernelFractional}
\begin{array}{lll}
K_0^{(\alpha)}(x,t)=\\ \displaystyle =\frac{1}{(2\pi)^{\frac{n}{2}}}\int_{Q_h}{(d_h(\xi)^2+m^2)^{\alpha}}\cosh\left(tL^{-1}\left(i\sqrt{d_h(\xi)^2+m^2}\right)\right)e^{-ix\cdot \xi}d\xi \\ \ \\ 
K_1^{(\alpha)}(x,t)=\\
\displaystyle = \frac{1}{(2\pi)^{\frac{n}{2}}}\int_{Q_h}{(d_h(\xi)^2+m^2)^{\alpha}}\dfrac{\sinh\left(tL^{-1}\left(i\sqrt{d_h(\xi)^2+m^2}\right)\right)}{i\sqrt{d_h(\xi)^2+m^2}}e^{-ix\cdot \xi}d\xi.
\end{array}
\end{eqnarray}

That corresponds to statement {\bf (i)} of {\bf Theorem \ref{mainResultdiffKleinGordonFractional}}.
Moreover, with aid of the lattice discretization
 \begin{eqnarray}
\label{RieszOperatorLattice}
\mathcal{R}_{h,\alpha}=\left(D_{h,\alpha}-m\gamma\right)(-\Delta_h+m^2)^{-\alpha}
\end{eqnarray}
of the \textit{Riesz type operator} $\displaystyle (D-m\gamma)(-\Delta+m^2)^{-\alpha}$ on $\BR_{h,\alpha}^n$ (cf.~\cite{Bersntein16}), we are able to recover the solution of the Dirac equation (\ref{diffDirac}). Such construction corresponds essentially to statements {\bf (iii)} and {\bf (iv)} of {\bf Theorem \ref{mainResultdiffKleinGordonFractional}}.

\begin{theorem}\label{mainResultdiffKleinGordonFractional}
Let $\mathcal{R}_{h,\alpha}$ be the discretized Riesz transform defined by eq. (\ref{RieszOperatorLattice}), and $K_0^{(\alpha)}$ and $K_1^{(\alpha)}$ the kernel functions defined through the integral equations (\ref{WaveKernelFractional}). Under the condition that $\Phi_0$ and $\Phi_1$ belong to $\mathcal{S}(\BR_{h,\alpha}^n;\BC \otimes \cl_{n,n})$, we have the following:
\begin{enumerate}
\item[{\bf (i)}] 
The ansatz function $\Psi(x,t)$ defined by means of the discrete convolution formula
\begin{eqnarray}
\label{FhFractionalKleinGordon}
\begin{array}{lll}
	\Psi(x,t)&=&\displaystyle \sum_{y \in \BR_{h,\alpha}^n}h^n (-\Delta_h+m^2)^{-\alpha}\Phi_0(y) ~K^{(\alpha)}_0(x-y,t)\\ \ \\
&+&\displaystyle \sum_{y \in \BR_{h,\alpha}^n}h^n (-\Delta_h+m^2)^{-\alpha}\Phi_1(y) ~K^{(\alpha)}_1(x-y,t)
\end{array}
\end{eqnarray}
solves the discretized Klein-Gordon equation (\ref{diffKleinGordon}).
\item[{\bf (ii)}] 
The ansatz function (\ref{FhFractionalKleinGordon}) solves the discretized Dirac equation (\ref{diffDirac}) whenever $$(-\Delta_h+m^2)^{-\alpha}\Phi_1(x)=i\mathcal{R}_{h,\alpha}\Phi_0(x).$$
\item [{\bf (iii)}] The inverse of the Riesz type operator $\mathcal{R}_{h,\alpha}$ is given by $$\left(\mathcal{R}_{h,\alpha}\right)^{-1}=(D_{h,\alpha}-m\gamma)(-\Delta_{h}+m^2)^{\alpha-1}.$$
\item[{\bf (iv)}] If $\Psi_0(x,t)={\bf P}_t[\Phi_0(x)]$ and 
\begin{center}
$\Psi_1(x,t)={\bf P}_t[(D_{h,\alpha}-m\gamma)(-\Delta_h+m^2)^{-1}\Phi_1(x)]$
\end{center}
 are two independent solutions of the discretized Dirac equation (\ref{diffDirac}), generated by the discrete convolution operator
\begin{eqnarray}
\label{FhFractionalDirac}
\begin{array}{lll}
{\bf P}_t[\Phi(x)]&=&\displaystyle \sum_{y \in \BR_{h,\alpha}^n}h^n (-\Delta_h+m^2)^{-\alpha}\Phi(y) ~K^{(\alpha)}_0(x-y,t)\\
&+&\displaystyle \sum_{y \in \BR_{h,\alpha}^n}h^n~ i\mathcal{R}_{h,\alpha}\Phi(y) ~K^{(\alpha)}_1(x-y,t),
\end{array}
\end{eqnarray}
then the function $$\Psi(x,t)=\frac{\Psi_0(x,t)+\Psi_0(x,-t)}{2}+\frac{\Psi_1(x,t)-\Psi_1(x,-t)}{2i}$$
solves the discretized Klein-Gordon equation (\ref{diffKleinGordon}).
\end{enumerate}
\end{theorem}

\proof
The proof of {\bf Theorem \ref{mainResultdiffKleinGordonFractional}} follows the same train of thought of the proof of {\bf Theorem \ref{mainResultdiffKleinGordon}} and {\bf Corollary \ref{mainResultdiffDirac}}. To avoid an overlap between the proof of these results we present only an abridged version of it, by sketching only the main ideas:

\vspace{0.20cm}
\textbf{Proof of (i):}

From the Laplace transform identity 
$$(d_h(\xi)^2+m^2)^{-\alpha}=\displaystyle \frac{1}{\Gamma(\alpha)}\int_0^{\infty} e^{-tm^2}e^{-td_h(\xi)^2}~t^\alpha \frac{dt}{t}$$
there holds the following identity for the operator $(-\Delta_h+m^2)^{-\alpha}$ defined through eq. (\ref{FractionalKleinGordon}): $$\mathcal{F}_{h,\alpha}[(-\Delta_h+m^2)^{-\alpha}\Phi](\xi)=(d_h(\xi)^2+m^2)^{-\alpha}\mathcal{F}_{h,\alpha}\Phi(\xi).$$ 

Therefore, the sequence of identities
\begin{eqnarray*}
	\mathcal{F}_{h,\alpha}\Psi(\xi,t)&=& \mathcal{F}_{h,\alpha}[(-\Delta_h+m^2)^{-\alpha}\Phi_0(\xi)]~\mathcal{F}_{h,\alpha}K_0^{(\alpha)}(\xi,t)+~\mathcal{F}_{h,\alpha}[(-\Delta_h+m^2)^{-\alpha}\Phi_1(\xi)] ~\mathcal{F}_{h,\alpha}K_1^{(\alpha)}(\xi,t)
\\ \ \\
&=&(d_h(\xi)^2+m^2)^{-\alpha}\mathcal{F}_{h,\alpha}\Phi_0(\xi)~\mathcal{F}_{h,\alpha}K_0^{(\alpha)}(\xi,t)+(d_h(\xi)^2+m^2)^{-\alpha}\mathcal{F}_{h,\alpha}\Phi_1(\xi)~\mathcal{F}_{h,\alpha}K_1^{(\alpha)}(\xi,t)\\ \ \\
	&=&\cosh\left(tL^{-1}\left(i\sqrt{d_h(\xi)^2+m^2}\right)\right)\mathcal{F}_{h,\alpha}\Phi_0(\xi)+ \dfrac{\sinh\left(tL^{-1}\left(i\sqrt{d_h(\xi)^2+m^2}\right)\right)}{i\sqrt{d_h(\xi)^2+m^2}}\mathcal{F}_{h,\alpha}\Phi_1(\xi)
\end{eqnarray*} 
yield straightforwardly from application of the \textit{discrete convolution property} (\ref{ConvolutionFhProperty}) underlying to the discrete Fourier transform $\mathcal{F}_{h,\alpha}$, and from the standard identities involving the wave kernels (\ref{WaveKernelFractional}):
\begin{eqnarray*}
	\mathcal{F}_{h,\alpha}K_0^{(\alpha)}(\xi,t)&=&(d_h(\xi)^2+m^2)^{\alpha}\cosh\left(tL^{-1}\left(i\sqrt{d_h(\xi)^2+m^2}\right)\right) \\
	\mathcal{F}_{h,\alpha}K_1^{(\alpha)}(\xi,t)&=&(d_h(\xi)^2+m^2)^{\alpha}\dfrac{\sinh\left(tL^{-1}\left(i\sqrt{d_h(\xi)^2+m^2}\right)\right)}{i\sqrt{d_h(\xi)^2+m^2}}.
\end{eqnarray*}

Thus $\mathcal{F}_{h,\alpha}\Psi(\xi,t)$ is a solution of the evolution problem (\ref{diffKleinGordonFourier}), and whence, the ansatz (\ref{FhFractionalKleinGordon}) solves the discretized Klein-Gordon equation (\ref{diffKleinGordon}).

\vspace{0.20cm}
\textbf{Proof of (ii):}

The proof that the condition $(-\Delta_h+m^2)^{-\alpha}\Phi_1(x)=i\mathcal{R}_{h,\alpha}\Phi_0(x)$ gives rise to the solution of the discretized Dirac equation (\ref{diffDirac}) is an immediate consequence of {\bf Corollary \ref{mainResultdiffDirac}} and of statement {\bf (i)} of {\bf Theorem \ref{mainResultdiffKleinGordonFractional}}.

\vspace{0.20cm}
\textbf{Proof of (iii):}

By noting that the Clifford vector ${\bf r}_{h,\alpha}(\xi)=({\bf z}_{h,\alpha}(\xi)-m\gamma)(d_h(\xi)^2+m^2)^{-\alpha}$
 corresponds to the Fourier multiplier of $\mathcal{F}_{h,\alpha}\circ \mathcal{R}_{h,\alpha} \circ \mathcal{F}_{h,\alpha}^{-1}$, and from the fact that the Clifford vector $${\bf s}_{h,\alpha}(\xi)=({\bf z}_{h,\alpha}(\xi)-m\gamma)(d_h(\xi)^2+m^2)^{\alpha-1}$$ is the inverse of ${\bf r}_{h,\alpha}(\xi)$, the proof that $\mathcal{S}_{h,\alpha}=(D_{h,\alpha}-m\gamma)(-\Delta_{h}+m^2)^{\alpha-1}$ equals to $\left(\mathcal{R}_{h,\alpha}\right)^{-1}$ follows straightforwardly from the set of identities
\begin{eqnarray*} \mathcal{F}_{h,\alpha}\left[\mathcal{S}_{h,\alpha}\mathcal{R}_{h,\alpha}\Phi\right](\xi)=&\mathcal{F}_{h,\alpha}\left[\mathcal{R}_{h,\alpha}\mathcal{S}_{h,\alpha}\Phi\right](\xi)=&\mathcal{F}_{h,\alpha}\Phi(\xi).
\end{eqnarray*}

\vspace{0.20cm}
\textbf{Proof of (iv):}

First, we recall that the splitting formulae
\begin{eqnarray*}
\begin{array}{lll}
\dfrac{{\bf P}_t[\Phi(x)]+{\bf P}_{-t}[\Phi(x)]}{2}&=&\displaystyle \sum_{y \in \BR_{h,\alpha}^n}h^n (-\Delta_h+m^2)^{-\alpha}\Phi(y) ~K^{(\alpha)}_0(x-y,t)\\ 
\dfrac{{\bf P}_t[\Phi(x)]-{\bf P}_{-t}[\Phi(x)]}{2i}&=&\displaystyle \sum_{y \in \BR_{h,\alpha}^n}h^n~ \mathcal{R}_{h,\alpha}\Phi(y) ~K^{(\alpha)}_1(x-y,t)
\end{array}
\end{eqnarray*}
yield from the parity properties involving the hyperbolic functions $\cosh$ (even function) and $\sinh$ (odd function).
By noting also that 
$$(D_{h,\alpha}-m\gamma)(-\Delta_h+m^2)^{-1}\Phi_1(y)=(-\Delta_h+m^2)^{-\alpha}\mathcal{R}_{h,1-\alpha}\Phi_1(x),$$
the proof that the function $$\Psi(x,t)=\frac{\Psi_0(x,t)+\Psi_0(x,-t)}{2}+\frac{\Psi_1(x,t)-\Psi_1(x,-t)}{2i}$$
provides a solution for the discretized Klein-Gordon equation (\ref{diffKleinGordon}) is rather immediate, since $\Psi(x,t)$ coincides with the ansatz (\ref{FhFractionalKleinGordon}).
\qed 

\begin{remark}[A Poisson semigroup counterpart]
Statement {\bf (iii)} of \textbf{Theorem \ref{mainResultdiffKleinGordonFractional}} may be seen as an hypercomplex analogue for the differential-difference Cauchy-Riemann equations, complementary to the one obtained in \cite[\textbf{Theorem 3.}]{CiaurriGRTV17} in terms of Poisson semigroup based representations.
In accordance with the discussion depicted in the end of subsection \ref{discreteHeatsemigroup}, we can
also see that the nonexistence of self-adjointness property for the discrete Dirac operators $D_{h,\alpha}$ (see \textbf{Remark \ref{LatticeDoublingGap}}) is not an obstacle to this approach.
\end{remark}

\section{Outlook of the main results}\label{Outlook}

This paper provides us a guideline to extend substancial part of the framework already done in \cite{BDattoliQ11,BaaskeBRS14,ConstalesDeRidder14,DattoliSGHP15,DattoliGHPS17,CiaurriGRTV17} to the differential-difference and to the difference-difference setting as well. To our best knowledge, there has been no literature paying attention to the description of the solutions of time-evolution problems as a blending between the continuous and the discrete side.

To mimic the construction depicted on \cite[\textsc{Chapter 2}]{Tao06} (see 
\textbf{Theorem \ref{mainResultdiffKleinGordon}}, \textbf{Corollary \ref{mainResultdiffDirac}} and {\bf Theorem \ref{mainResultdiffKleinGordonFractional}}) we adopted in section \ref{Setting} the \textit{toroidal Fourier framework} proposed by Ruzhansky and Turunen on their book \cite[Part II, Chapter 3]{RuzhanskyT10} to exploit the framework proposed by G\"urlebeck and Spr\"ossig in \cite[subsection 5.2]{GuerlebeckSproessig97} to lattices of the type $\BR^n_{h,\alpha}=(1-\alpha)h\BZ^n+\alpha h\BZ^n$ ($h>0$ and $0<\alpha<\frac{1}{2}$), based on the one-to-one correspondence between the so-called $n-$\textit{Brioullin zone} $Q_h=\left(-\frac{\pi}{h},\frac{\pi}{h}\right]^n$ and the toroidal manifold $\BR^n/\frac{2\pi}{h}\BZ^n$ (cf.~\cite{Rabin82}).
We also propose in section \ref{Setting} a
\textit{fractional regularization} for discrete Dirac operators acting on the lattices $h\BZ^n$ $(\alpha\rightarrow 0)$ and $\frac{h}{2}\BZ^n$ $(\alpha\rightarrow \frac{1}{2})$, based on construction of a wide class of Fourier multipliers with values on the Clifford algebra with signature $(n,n)$.

From the results obtained in section \ref{FurtherApplications} (namely \textbf{Theorem \ref{mainResultdiffKleinGordonFractional}}) we believe that the proposed approach does not offer only a wise strategy to determine discrete counterparts for the results provided by the papers \cite{BDattoliQ11,BaaskeBRS14,ConstalesDeRidder14,DattoliSGHP15,DattoliGHPS17,CiaurriGRTV17}. In the shed of the fractional calculus formulation proposed recently by Bernstein (cf.~\cite{Bersntein16,Bersntein17}), it may also provides us a meaningful way to generalize the results of \cite{CKKS14,CKK15}, where only the properties of the \textit{discrete Fourier transform} depicted in \cite[subsection 5.2]{GuerlebeckSproessig97} were taken into account.

As a whole, the fractional integration approach combined with a \textit{space-time Fourier inversion type formula} has been revealed as an exceptionally well-suited tool to represent, in an operational way, the discrete convolution representations underlying to the solutions of an equation of Klein-Gordon type. As we have noticed on subsection \ref{discreteHeatsemigroup}, this is ultimately due to the possibility of describe the superposition of the wave-type propagators
\begin{center}
$\cosh(tL^{-1}(\Delta_h-m^2))$ resp. $\dfrac{\sinh(tL^{-1}(\sqrt{\Delta_h-m^2}))}{\sqrt{\Delta_h-m^2}}$
\end{center} in terms of its Fourier-Laplace multipliers endowed by the \textit{discrete heat kernel} $\exp\left(\frac{p\tau^2}{4\tau}\Delta_h\right)$ (see, for instance, \textbf{Remark \ref{LaplaceWeierstrassRemark}}). For a general overview of this framework, we refer to \cite[Chapter 5]{SamkoEtAl93}. 

{\sc Summing up:} The discrete Fourier transform framework brings into the representation of the solution for an evolution type equation over the momentum space $Q_h \times T$. In the future it can also be useful to look in depth for the associated hypersingular operator representations on the \textit{space-time toroidal manifold} $(\BR^n/\frac{2\pi}{h}\BZ^n) \times (\BR/\frac{4\pi}{\tau}\BZ)$ (see, for instance, \textbf{Remark \ref{discreteCKremark}}). Although the Fourier-Laplace type multipliers are a little trickier to compute, due to the fractional calculus technicalities, the Fourier modes $e^{-i(\omega t+x\cdot \xi)}$ associated to the space-time Fourier inversion formula are more easier to treat on the space of tempered distributions, in comparison to the description of fractional integro-differential operators on the manifold $(\BR^n/h\BZ^n) \times [0,\infty)$ (see, for instance, subsection 24.10 of \cite[Chapter 5]{SamkoEtAl93}).

\appendix
\section{The Exponential Generating Function Connection}\label{EGFsection}

In the paper \cite{FaustinoMonomiality14} we have considered the exponential generating functions (EGF) $\exp(tL^{-1}(s))$ to derive hypercomplex formulations for Appell sets and \textit{Exponential Generating Function} (EGF). 
With the aim of construct umbral counterparts for the wave type propagators
\begin{center}
 $\cosh\left(t\sqrt{\Delta_h-m^2}\right)$ resp. 
$\dfrac{\sinh\left(t\sqrt{\Delta_h-m^2}\right)}{\sqrt{\Delta_h-m^2}}$,
\end{center}
we consider here a wise adaptation of \cite[Corollary 1.1.15]{Faustino09} for $\cosh(tL^{-1}(s))$ and $\sinh(tL^{-1}(s))$. That corresponds to the following result:
\begin{theorem}\label{EGFcoshsinh}
	The formal series representation of $\cosh(tL^{-1}(s))$ and $\sinh(tL^{-1}(s))$
	determined by the \textit{delta operator} $L_t=L(\partial_t)$ are given by
	\begin{eqnarray*}
		\cosh(tL^{-1}(s))=\sum_{k=0}^\infty \dfrac{m_{2k}(t)}{(2k)!}s^{2k} & \mbox{and}& \displaystyle
		\sinh(tL^{-1}(s))=\sum_{k=0}^\infty \dfrac{m_{2k+1}(t)}{(2k+1)!}s^{2k+1}, 
	\end{eqnarray*}
	where $\{m_k(t)~:~k \in\BN_0\}$ is a basic polynomial sequence associated to $L_t$.
\end{theorem}

\textbf{Proof of Theorem \ref{EGFcoshsinh}}
First, we recall that from \cite[Corollary 1.1.15]{Faustino09}, the exponentiation operator 
$\exp(s\partial_t)$ may be formally represented as 
$$
\exp(t\partial_s)=\sum_{k=0}^\infty \frac{m_k(t)}{k!}L(\partial_s)^k.
$$

Then, we have
\begin{eqnarray*}
	\cosh(t\partial_s)&=&\frac{\exp(t\partial_s)+\exp(-t\partial_s)}{2}=\sum_{k=0}^\infty \frac{m_{2k}(t)}{(2k)!}L(\partial_s)^{2k} \\
	\sinh(t\partial_s)&=&\frac{\exp(t\partial_s)-\exp(-t\partial_s)}{2}=\sum_{k=0}^\infty \frac{m_{2k+1}(t)}{(2k+1)!}L(\partial_s)^{2k+1}.
\end{eqnarray*}

By applying the isomorphism theorem (cf.~\cite[Theorem 2.2.1]{Roman84}) one get a one-to-one correspondence between $\cosh(t\partial_s)$ resp. $\sinh(t\partial_s)$ with the formal power series expansions
\begin{eqnarray}
\label{coshsinhIdentities}
\begin{array}{lll}
\cosh(ts)&=&\displaystyle \sum_{k=0}^\infty \frac{m_{2k}(t)}{(2k)!}L(s)^{2k} \\ \ \\
\displaystyle \sinh(ts)&=&\displaystyle \sum_{k=0}^\infty \frac{m_{2k+1}(t)}{(2k+1)!}L(s)^{2k+1}.
\end{array}
\end{eqnarray}

The conclusion of {\bf Theorem \ref{EGFcoshsinh}} for $\cosh(tL^{-1}(s))$ and $\sinh(tL^{-1}(s))$ then follows from the substitution $s\rightarrow L^{-1}(s)$ on both sides of  (\ref{coshsinhIdentities}).

\qed

Let us now take a close look for the \textit{Exponential Generating Functions} (EGF) $\G(\s,t)$ of hypercomplex type defined by means of equation (\ref{EGFLt}). 

It is quite easy to see that the condition $m_0(t)=1$ and the lowering properties $L_tm_k(t)=m_{k-1}(t)$ $(k\in \BN)$ lead us naturally to the condition $\G(\s,0)=1$, and to the eigenvalue property
\begin{eqnarray}
\label{EGFLtEigenvalue}L_t\G(\s,t)=\s\G(\s,t).
\end{eqnarray}

Also, it is worth stressing that {\bf Theorem \ref{EGFcoshsinh}} may me extended/generalized for hypercomplex variables. In particular, \textbf{Theorem \ref{EGFhypercomplex}} corresponds to a wise generalization of {\bf Theorem \ref{EGFcoshsinh}}. The proof proceeds as follows:

\textbf{Proof of Theorem \ref{EGFhypercomplex}:}
First, we recall that 
the even resp. odd part of $\G(\s,t)$ may be expressed as
\begin{eqnarray*}
	\frac{\G(\s,t)+\G(-\s,t)}{2}&=&\sum_{k=0}^{\infty} \dfrac{m_{2k}(t)}{(2k)!}\s^{2k} \\
	\frac{\G(\s,t)-\G(-\s,t)}{2}&=&\sum_{k=0}^{\infty} \frac{m_{2k+1}(t)}{(2k+1)!}\s^{2k+1}.
\end{eqnarray*} 

Finally, from direct application of {\bf Theorem \ref{EGFcoshsinh}}, we recognize that for the Clifford numbers of the form 
\begin{eqnarray*}
	\s=re^{i\phi}\omega, & \mbox{with}& r\geq 0,~~ -\pi<\phi\leq \pi ~~\&~~ \omega^2=+1
\end{eqnarray*}
the above set of identities equals to
\begin{eqnarray*}
	\dfrac{\G(re^{i\phi}\omega,t)+\G(-re^{i\phi}\omega,t)}{2}&=&\cosh(tL^{-1}\left(re^{i\phi}\right)) \\ \dfrac{\G(re^{i\phi}\omega,t)-\G(-re^{i\phi}\omega,t)}{2}&=&\omega\sinh(tL^{-1}\left(re^{i\phi}\right)),
\end{eqnarray*}
since $\s^{2k}=\left(re^{i\phi}\right)^{2k}\left(\omega^{2}\right)^k=\left(re^{i\phi}\right)^{2k}$ and $\s^{2k+1}=\s \s^{2k}=\left(re^{i\phi}\right)^{2k+1}\omega$, concluding in this way that 
$$ 
\G(re^{i\phi}\omega,t)=\cosh\left(tL^{-1}\left(re^{i\phi}\right)\right)+\omega\sinh\left(tL^{-1}\left(re^{i\phi}\right)\right),
$$
as desired.
\qed



\newpage

\begin{thebibliography}{1}



\bibitem{BaaskeBRS14}
{Baaske, F., Bernstein, S., De Ridder, H., \& Sommen, F. (2014). \textit{On solutions of a discretized heat equation in discrete Clifford analysis}. Journal of Difference Equations and Applications, 20(2), 271-295.}

\bibitem{BDattoliQ11}
{Babusci, D., Dattoli, G., \& Quattromini, M. (2011). \textit{Relativistic equations with fractional and pseudodifferential operators}. Physical Review A, 83(6), 062109.}

\bibitem{Bersntein16}
Bernstein, S. (2016). \textit{A Fractional Dirac Operator}. In Noncommutative Analysis, Operator Theory and Applications (pp. 27-41). Birkh\"auser, Cham.

\bibitem{Bersntein17}
Bernstein, S. (2017). \textit{Fractional Riesz-Hilbert-Type Transforms and Associated Monogenic Signals}. Complex Analysis and Operator Theory, 11(5), 995-1015.

\bibitem{CKKS14}
Cerejeiras, P., K\"ahler, U., Ku, M., \& Sommen, F. (2014). \textit{Discrete hardy spaces}. Journal of Fourier Analysis and Applications, 20(4), 715-750.

\bibitem{CKK15}
Cerejeiras, P., K\"ahler, U., \& Ku, M. (2015). \textit{Discrete Hilbert boundary value problems on half lattices}. Journal of Difference Equations and Applications, 21(12), 1277-1304.


\bibitem{CiaurriGRTV17}Ciaurri, \'O., Gillespie, T. A., Roncal, L., Torrea, J. L., \& Varona, J. L. (2017). \textit{Harmonic analysis associated with a discrete Laplacian}. Journal d'Analyse Math\'ematique, 132(1), 109-131.

\bibitem{ConstalesDeRidder14}
{Constales, D., \& De Ridder, H. (2014). \textit{A Compact Cauchy-Kovalevskaya Extension Formula in Discrete Clifford Analysis}. Advances in Applied Clifford Algebras, 24(4), 1005-1010.}


\bibitem{Dattoli97}
Dattoli, G., Ottaviani, P. L., Torre, A., \& Vázquez, L. (1997). \textit{Evolution operator equations: Integration with algebraic and finite difference methods. Applications to physical problems in classical and quantum mechanics and quantum field theory}. La Rivista del Nuovo Cimento (1978-1999), 20(2), 3.

\bibitem{DattoliSGHP15}{Dattoli, G., Sabia, E., G\'orska, K., Horzela, A., \& Penson, K. A. (2015). \textit{Relativistic wave equations: an operational approach}. Journal of Physics A: Mathematical and Theoretical, 48(12), 125203.}

\bibitem{DattolliT15}
{Dattoli, G., \& Torre, A. (2015). \textit{Root operators and ``evolution equations''}. Mathematics, 3(3), 690-726.}

\bibitem{DattoliGHPS17}
{Dattoli, G., K. G\'orska, A. Horzela, K. A. Penson, and E. Sabia. \textit{Theory of relativistic heat polynomials and one-sided L\'evy distributions}. Journal of Mathematical Physics 58, no. 6 (2017): 063510.}


\bibitem{FaustinoGHK06}{Faustino, N., G\"urlebeck, K., Hommel, A., \& K\"ahler, U. (2006). \textit{Difference potentials for the Navier-Stokes equations in unbounded domains}. Journal of Difference Equations and Applications, 12(6), 577-595.}

\bibitem{CFaustinoV08}{Cerejeiras, P., Faustino, N., \& Vieira, N. (2008). \textit{Numerical Clifford analysis for nonlinear Schr\"odinger problem}. Numerical Methods for Partial Differential Equations, 24(4), 1181-1202.}


\bibitem{Faustino09}
Faustino, N. J. R. (2009). \textit{Discrete Clifford analysis} (Doctoral dissertation, Universidade de Aveiro (Portugal)),
ix+130 pages.


\bibitem{Faustino10}
{Faustino, N. (2010). \textit{Further results in discrete Clifford analysis}. In Progress in Analysis and Its Applications (pp. 205-211).}

\bibitem{FaustinoR11}{Faustino, N., \& Ren, G. (2011). \textit{(Discrete) Almansi type decompositions: an umbral calculus framework based on $\mathfrak{osp}(1| 2)$ symmetries}. Mathematical Methods in the Applied Sciences, 34(16), 1961-1979.}


\bibitem{FaustinoMonomiality14}
Faustino, N. (2014). \textit{Classes of hypercomplex polynomials of discrete variable based on the quasi-monomiality principle}. Applied Mathematics and Computation, 247, 607-622.

\bibitem{FaustinoKGordonDirac16}
Faustino, N. (2016). \textit{Solutions for the Klein-Gordon and Dirac equations on the lattice based on Chebyshev polynomials}. Complex Analysis and Operator Theory, 10(2), 379-399.

\bibitem{FaustinoMMAS17}
Faustino, R., \& Jos\'e, N. (2017). \textit{A conformal group approach to the Dirac-K\"ahler system on the lattice}. Mathematical Methods in the Applied Sciences, 40(11), 4118-4127.


\bibitem{FaustinoBayesian17}
Faustino, Nelson. \textit{Hypercomplex Fock States for Discrete Electromagnetic Schr\"odinger Operators: A Bayesian Probability Perspective}. Applied Mathematics and Computation 315 (2017): 531-548.

\bibitem{GuerlebeckSproessig97}{G\"urlebeck, K., \& Spr\"ossig, W. (1997). \textit{Quaternionic and Clifford calculus for physicists and engineers}. Wiley.}


\bibitem{MasonChebyshev93}{Mason, J. C. (1993). \textit{Chebyshev polynomials of the second, third and fourth kinds in approximation, indefinite integration, and integral transforms}. Journal of Computational and Applied Mathematics, 49(1-3), 169-178.}


\bibitem{NN81}{Nielsen, H. B., \& Ninomiya, M. (1981). \textit{A no-go theorem for regularizing chiral fermions}. Physics Letters B, 105(2-3), 219-223.}



\bibitem{Rabin82}
Rabin, J. M. (1982). \textit{Homology theory of lattice fermion doubling}. Nuclear Physics B, 201(2), 315-332.

\bibitem{Roman84}
{Roman S. (1984). \textit{The Umbral Calculus}. Academic Press.}

\bibitem{RuzhanskyT10}
Ruzhansky, M., \& Turunen, V. (2010). \textit{Pseudo-differential operators and symmetries: background analysis and advanced topics (Vol. 2)}. Springer Science \& Business Media.


\bibitem{SamkoEtAl93}
Samko, S. G., Kilbas, A. A., \& Marichev, O. I. (1993). \textit{Fractional integrals and derivatives. Theory and Applications}, Gordon and Breach, Yverdon, 1993, 44.


\bibitem{SaxenaEtAl02}
{Saxena, R. K., Mathai, A. M., \& Haubold, H. J. (2002). \textit{On fractional kinetic equations}. Astrophysics and Space Science, 282(1), 281-287.}

\bibitem{Tao06}
Tao, T. (2006). \textit{Nonlinear dispersive equations: local and global analysis} (No. 106). American Mathematical Soc..

\bibitem{VazRoldao16}{
Vaz Jr, J., \& da Rocha Jr, R. (2016). \textit{An introduction to Clifford algebras and spinors}. Oxford University Press.}



\end{thebibliography}
\end{document}